\documentclass[a4paper]{article}
\pdfoutput=1
\usepackage{jcappub}
\usepackage{bbold}
\usepackage{mathtools}
\usepackage[normalem]{ulem}
\usepackage{xspace}
\usepackage{float}
\usepackage[makeroom]{cancel}
\usepackage{afterpage}
\usepackage{booktabs}
\usepackage{multirow}
\usepackage{enumerate}

\def\laq{~\raise 0.4ex\hbox{$<$}\kern -0.8em\lower 0.62ex\hbox{$\sim$}~}
\def\gaq{~\raise 0.4ex\hbox{$>$}\kern -0.7em\lower 0.62ex\hbox{$\sim$}~}

\newcommand{\class}{{\sc class}\xspace}
\newcommand{\classgal}{{\sc class}gal\xspace}

\def \ra {\rightarrow}

\def \la {\lambda}

\def \La {\Lambda}
\def \De {\Delta}
\def \de {\delta}

\def \Om {\Omega}

\def\laq{~\raise 0.4ex\hbox{$<$}\kern -0.8em\lower 0.62ex\hbox{$\sim$}~}
\def\gaq{~\raise 0.4ex\hbox{$>$}\kern -0.7em\lower 0.62ex\hbox{$\sim$}~}

\def\ch#1{#1}

\def \ra {\rightarrow}

\def \La {\Lambda}

\def \Om {\Omega}

\newcommand{\bn}{{\bf n}}

\newcommand{\HH}{\mathcal H}
\newcommand{\OO}{\mathcal O}

\newcommand{\bx}{\boldsymbol{x}}

\def\laq{~\raise 0.4ex\hbox{$<$}\kern -0.8em\lower 0.62ex\hbox{$\sim$}~}
\def\gaq{~\raise 0.4ex\hbox{$>$}\kern -0.7em\lower 0.62ex\hbox{$\sim$}~}

\def\be{\begin{equation}}
\def\ee{\end{equation}}
\def\bea{\begin{eqnarray}}
\def\eea{\end{eqnarray}}
\def\bean{\begin{eqnarray*}}
\def\eean{\end{eqnarray*}}

\newcommand{\overbar}[1]{\mkern 1.5mu\overline{\mkern-1.5mu#1\mkern-1.5mu}\mkern 1.5mu}

\newcommand{\myvec}[1]{\boldsymbol{#1}}
\newcommand{\uvec}[1]{\myvec{\hat{#1}}}
\newcommand{\nvec}{\uvec{n}}

\newcommand{\dipole}{\myvec{d}}
\newcommand{\dint}{\dipole_\mathrm{int}}
\newcommand{\dkin}{\dipole_\mathrm{kin}}

\newcommand{\dipoleNC}{\myvec{d}^{N}}
\newcommand{\dkinNC}{\dipole_\mathrm{kin}^{N}}
\newcommand{\dsnNC}{\dipole_\mathrm{SN}^{N}}

\newcommand{\dipoleW}{\myvec{d}^W}
\newcommand{\dkinW}{\dipole_\mathrm{kin}^W}
\newcommand{\dsnW}{\dipole_\mathrm{SN}^W}

\newcommand{\vecbeta}{\myvec{\beta}}

\newcommand{\ndens}{\rho}
\newcommand{\ndensW}{\ndens_W}
\newcommand{\ndensWbar}{\bar{ \ndens}_W}

\newcommand{\Flim}{F_\mathrm{min}}
\newcommand{\order}[1]{\OO(#1)}
\newcommand{\defined}{\equiv}

\newcommand{\rframe}[1]{#1_\mathrm{rest}}

\newcommand{\gaussian}{\mathcal{N}}
\newcommand{\fsky}{f_\mathrm{sky}}
\newcommand{\Ntot}{N_\mathrm{tot}}
\newcommand{\ndensbar}{\bar{\ndens}}
\newcommand{\nbar}{\overbar{n}}

\newcommand{\integral}[3]{\int_{#2}^{#3} \mathrm{d} #1 \, }

\newcommand{\amp}{B}
\newcommand{\ampW}{\amp^W}
\newcommand{\ampNC}{\amp^{N}}

\newcommand{\trecs}{T-RECS\xspace}

\newcommand{\healpy}{healpy\xspace}

\newcommand{\healpix}{HEALPix\xspace}
\newcommand{\fitdipole}{{\tt fit\_dipole}\xspace}

\newcommand{\degree}{^\circ}
\newcommand{\arcsec}{\mathrm{arcsec}}

\newcommand{\expect}[1]{\langle #1 \rangle }

\newcommand{\estNC}{\myvec{d}_\mathrm{est}^{N}}
\newcommand{\estW}{\myvec{d}_\mathrm{est}^W}
\newcommand{\estBeta}{\myvec{\beta}^\mathrm{est}}
\newcommand{\estInt}{\myvec{d}_\mathrm{int}^\mathrm{est}}

\newcommand{\figref}[1]{Fig.~\ref{#1}}
\newcommand{\secref}[1]{Section~\ref{#1}}

\newcommand{\OPTindex}{\alpha_{\mathrm{opt}}}
\newcommand{\SKAindex}{\alpha_{\mathrm{SKA}}}

\newcommand{\dsigW}{\Delta_W}

\newcommand{\CbetaSN}{C_\beta^\mathrm{SN}}
\newcommand{\CintSN}{C_\mathrm{int}^\mathrm{SN}}

\newcommand{\dip}{d^N}

\newcommand{\fevo}{f_\mathrm{evo}}
\newcommand{\Ncurl}{\mathcal{N}}
\newcommand{\der}[2]{\frac{\mathrm{d} #1}{\mathrm{d} #2}}
\newcommand{\del}{\partial}
\newcommand{\partialder}[2]{\frac{\del #1}{\del #2} }

\title{A new \ch{way to test} the Cosmological Principle: measuring our peculiar velocity and the large-scale anisotropy independently}
\author{Tobias Nadolny, Ruth Durrer, Martin Kunz and Hamsa Padmanabhan}

\affiliation{
Universit\'e de Gen\`eve, D\'epartement de Physique Th\'eorique and CAP,
24 quai Ernest-Ansermet, CH-1211 Gen\`eve 4, Switzerland
}

\emailAdd{tobias.nadolny@gmx.de}
\emailAdd{ruth.durrer@unige.ch}
\emailAdd{martin.kunz@unige.ch}
\emailAdd{hamsa.padmanabhan@unige.ch}

\date{June 2021}

\abstract{
We present a novel approach to disentangle two key contributions to the largest-scale anisotropy of the galaxy distribution: (i) the intrinsic dipole due to clustering and anisotropic geometry, and (ii) the kinematic dipole due to our peculiar velocity. Including the redshift and angular size of galaxies, in addition to their fluxes and positions allows us to measure both the direction and amplitude of our velocity  independently of the intrinsic dipole of the source distribution.
We find that this new approach applied to future galaxy surveys (LSST and Euclid) and a SKA radio continuum survey \ch{will allow} to measure our velocity ($\beta = v/c$) with a relative error in the amplitude $\sigma(\beta)/\beta \sim (1.3 - 4.5)\%$  and in direction, $\theta_{\beta} \sim 0.9\degree - 3.9\degree$, well beyond what can be achieved 
when analysing only the number count dipole.
We also find that galaxy surveys are able to measure the intrinsic large-scale anisotropy with a relative uncertainty of $\lesssim 5\%$ (\ch{measurement error, not including} cosmic variance).
Our method enables two simultaneous tests of the Cosmological Principle: comparing the observations of our peculiar velocity with the CMB dipole, and testing for a significant intrinsic anisotropy on large scales which \ch{would indicate} effects beyond the standard cosmological model.
}

\begin{document}

\maketitle

\section{Introduction}
The dipole of the cosmic microwave background (CMB) is about a hundred times larger than the higher multipoles. The simplest interpretation of this observation is to assume that the solar system barycenter moves with a velocity of  $v_{\rm CMB}\simeq   369.82 \pm 0.114$ km/s towards $(l, b) = (264.021 \degree \pm 0.011 \degree, 48.253 \degree \pm 0.005 \degree)$
 with respect to the surface of last scattering of the CMB photons~\cite{pdg2019,2020A&A...641A...1P}. This velocity has not only been measured with high accuracy, but the Doppler effect has also tentatively been confirmed by its second order effects (aberration and modulation) on higher multipoles~\cite{Aghanim:2013suk,PhysRevLett.127.101301,2021_Saha}.
 Furthermore, the motion of the earth around the sun is seen as a variation of the CMB dipole~\cite{1992ApJ...391..466B}. If the cosmic structures in the Universe grew out of small initial fluctuations in a homogeneous and isotropic spacetime by gravitational instability, the CMB rest frame should be shared by the rest
 frame of the galaxy distribution on sufficiently large scales up to small fluctuations.
 Therefore, our velocity (the velocity of the solar system barycenter) with respect to the mean rest frame of distant galaxies is expected to match $v_\mathrm{CMB}$.
 From this, one can predict the amplitude and direction of a kinematic dipole in the number density of galaxies~\cite{1984MNRAS.206..377E}.
 
Since the discovery of the CMB dipole in 1969~\cite{1969Natur.222..971C,1971Natur.231..516H} researchers in the field have tried to determine the dipole of the galaxy distribution to test the effect of our motion and isotropy. A first significant step in this direction was~Ref. \cite{1994ApJ...425..418L} which tried to determine the velocity of the local group with respect to far away galaxy clusters (Abell clusters). 
Many more analyses have followed, most notably those determining the dipole of very large radio surveys like the NVSS~\cite{nvss} containing more than one million radio galaxies, see e.g.~\cite{Blake:2002gx,Rubart:2013tx,Bengaly:2017slg,Tiwari:2018hrs,Singal:2011,Gibelyou:2012,Tiwari:2015_NVSS,Tiwari:2016_Cluster_Simulation,Colin:2017_NVSSUMSS,Tiwari:2019_TGSS} and Ref.~\cite{siewert:2020_quad_est} where all previous results are collected in an appendix.

Even though these papers do not all find consistent results, they typically obtain a dipole which points roughly in the right direction but is significantly larger in magnitude than the expected kinematic dipole.
A similar analysis using quasars has prompted some researchers in the field to doubt the homogeneity and isotropy of the Universe at large scales~\cite{Secrest_2021}.
In this paper, we revisit the dipole of the galaxy distribution. In addition to the dipole of their number counts, we also investigate the dipole in the distribution of the galaxy fluxes, sizes and redshifts.
We show how an optimal combination of two such dipole measurements allows us to isolate both, the intrinsic dipole and the kinematic dipole. The idea to separate these two contributions was first explored in \cite{Tiwari:2015_NVSS} (see also \cite{Singal_2019} for a related approach).
Perhaps most importantly, if the next generation large-scale structure surveys confirm the excess in dipole amplitude, this novel approach has the potential to distinguish whether the excess comes from an intrinsic anisotropy of our Universe, or because our velocity relative to the matter rest frame differs from the CMB velocity. It thus provides two important tests of the standard cosmological paradigm.

The paper is organized as follows. We first briefly discuss the earlier analyses of the dipole in source number counts mentioned above, as well as  the effects of shot noise on these results in Section~\ref{s:rev}. In Section~\ref{s:new} we then present our new measurement method: we show that by measuring both, the flux and the size, or the flux and the redshift of sources with sufficient accuracy, one can in principle disentangle an intrinsic dipole due to clustering or an anisotropic universe from the kinematic dipole due to our motion. 
In Section~\ref{s:applications} we study how this new method can be applied to future experiments, by using an optical survey with redshift measurements like the LSST or Euclid, and by using the most advanced radio observatory, the Square Kilometre Array (SKA). We summarize our conclusions in Section~\ref{s:con}.
\vspace{0.2cm}

{\bf Notation:} Throughout this paper we set the speed of light $c=1$ so that
$v_{\rm CMB}=\beta \simeq 1.234\times10^{-3}$.  Spatial 3D vectors are indicated by boldface letters and the unit vector by an additional hat, e.g.,~$\vecbeta = \beta \uvec{\beta}$.

\section{Review of the number count dipole}\label{s:rev}
The dipole in the number count density \ch{$\ndens(>\Flim, \uvec{n}) \defined {\mathrm{d} N}/{\mathrm{d} \Omega} (>\Flim,\uvec{n})$} of galaxies with flux larger than $\Flim$ that we observe in a direction $\uvec{n}$ is defined by
\be
\ndens (>\Flim, \uvec{n}) = ( 1 + \dipoleNC \cdot \uvec{n} ) \ndensbar +\OO(\hat{n}_j^2)
\, ,
\label{eq:dipole_definition}
\ee
where $\ndensbar$ is the average number density and $\OO(\hat{n}_j^2)$ denotes higher multipoles. In what follows, we generically refer to galaxies as `sources', neglecting differences between their types.

\subsection{Contributions}
\label{s:EB_derivation}

There are three contributions to the number count dipole
\be
\dipoleNC  = \dkinNC + \dint + \dsnNC
\, .
\label{eq:combined_dipole}
\ee
The superscript $N$ stands for `number count dipole' and will be relevant in Section~\ref{s:new} where we also consider the dipole of other observables.
The first contribution in \eqref{eq:combined_dipole} is the kinematic dipole due to the observer's motion with respect to the mean rest frame of the sources. It is given by \cite{1984MNRAS.206..377E}
\be
\dkinNC = [2 + x(1+\alpha)] \vecbeta \,.
\label{eq:EllisBaldwinResult}
\ee
Here, it is assumed that both the cumulative number of sources as a function of the flux cut \ch{$N(>\Flim)$} and the frequency spectrum of the flux obey power laws, such that \ch{$N(>\Flim) \propto \Flim^{-x}$} in the vicinity of the flux limit $F_\mathrm{min}$ and $F \propto \nu^{-\alpha}$ \ch{in the observed frequency range}.
Typical values for a radio survey are $x \approx 1$ and $\alpha \approx 0.75$ \ch{(see e.g.~\cite{Blake:2002gx})}.
\ch{These values are empirical and are not derived by any theoretical consideration.}
The radio galaxy \ch{frequency} spectrum is observationally well-characterized by a power law with $\alpha \sim 0.8$, e.g.~\cite{kellerman1988}. If the flux of a source is observed at different frequencies, $\alpha$ can be estimated from the survey data.
\ch{If not, other observations of the frequency spectrum need to be invoked for an estimate of $\alpha$.}
The slope of the source count distribution is found to flatten below a value of $x= 1.5$ at fainter levels, motivating $x \sim 1$ at the limiting flux, e.g.~\cite{kellermann1987, white1997,Bengaly:2017slg}.
\ch{In fact, the value of $x$ can 
be determined directly from a given survey. For this one has to
fit a power law to the cumulative source counts, $N(>\Flim)\propto \Flim^{-x}$, close to the lower flux limit, $F_{\min}$.}

The first term in \eqref{eq:EllisBaldwinResult} comes from the change of the solid angle in direction $\uvec{n}$ perceived by a moving observer, which is given by 
\be
{\mathrm{d} \Omega}/{\mathrm{d} \rframe{\Omega}}
= \delta(\uvec{n})^{-2}
= (1 - 2\vecbeta \cdot  \uvec{n}) +\order{\beta^2} \, , \mbox{ where }~~  \delta(\uvec{n}) = (1+\uvec{n} \cdot \vecbeta)/\sqrt{1-\beta^2}\,.
\label{eq:solid_angle_change}
\ee
This is the special relativistic aberration, i.e., the shift of the observed direction towards the direction of motion, given by
$
\uvec{n} \cdot  \uvec{\beta} = 
(\beta + \rframe{\uvec{n}} \cdot  \uvec{\beta})/(1 + \rframe{\uvec{n}} \cdot  \vecbeta)$~\cite{Einstein_1905}.
The terms proportional to $x$ and to $x\alpha$ in \eqref{eq:EllisBaldwinResult} are due to the Doppler boost of the flux and the Doppler shift of the frequency $\nu=\rframe{\nu}\delta(\uvec{n})$, respectively.
Combining these effects, the received flux is $F=\rframe{F}\delta (\uvec{n}) ^{(1+\alpha)}$ when observing at a fixed frequency.
Therefore, the number counts above a limiting flux threshold are changed by $\delta(\uvec{n})^{x(1+\alpha)} \approx 1+x(1+\alpha)\vecbeta \cdot \uvec{n} $.
Adding these terms together, the observed number count density is 
\be\label{e:EllisBald}
\ndens (>\Flim,\nvec)  =  \ndens_\mathrm{rest} (\nvec) \Big(1+[2+x(1+\alpha)]\vecbeta \cdot \uvec{n} \Big) \, ,
\ee
leading to the result for the kinematic dipole given in \eqref{eq:EllisBaldwinResult}.
With $x\approx 1$ and $\alpha\approx 0.75$, the kinematic dipole is about $4.6 \times 10^{-3}$.
A discussion of the kinematic dipole in number counts as a function of redshift is also given in~\cite{Maartens:2017qoa}.  \ch{In Appendix~\ref{app:gr_kinematic} we show that the result derived in~\cite{Maartens:2017qoa} is equivalent to \eqref{e:EllisBald}.}

The second contribution is the intrinsic dipole $\dint$, which is the largest-scale anisotropy in the rest frame distribution \ch{$\rframe{\ndens}(\nvec)$}. In a statistically isotropic universe it is due to the fact that even if the observer is perfectly at rest with respect to the background cosmology, the galaxies have peculiar motions and density fluctuations which lead to a dipole in their distribution.
This dipole is therefore often called the `clustering dipole' or `structure dipole'. \ch{It has contributions from density fluctuations, redshift space distortions, gravitational lensing and other relativistic terms. We discuss it in more detail in Appendix~\ref{app:Dip-int}.}
If the Universe does not obey the Cosmological Principle,
but has significant large scale anisotropies, e.g.\ in a Bianchi model, this intrinsic dipole can be significantly larger.

\begin{figure}
    \centering
    \includegraphics[width=0.6\textwidth]{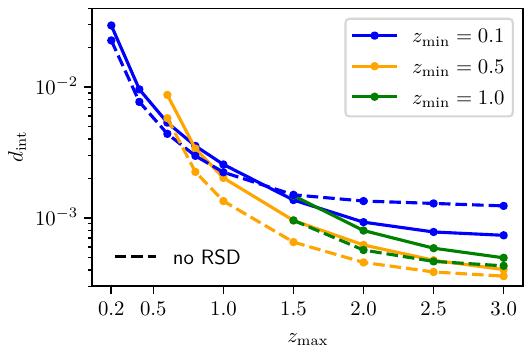}
    \caption{The intrinsic dipole $d_\mathrm{int} = \sqrt{9 C_1^\mathrm{int} /({4\pi})}$ of galaxy number counts expected in the standard $\Lambda$CDM cosmological model. A tanh-smoothed tophat window with different redshift limits $z_\mathrm{min}$ and $z_\mathrm{max}$ is used for the calculation with \classgal. The magnification bias is set to $s=0$ and the bias to $b =1.5$. The dashed lines show the intrinsic dipole without the presence of redshift space distortions.
    For comparison, the kinematic dipole amplitude is approximately $4.6 \times 10^{-3}$. 
    \label{f:intrinsic}
    }
\end{figure}

Within the cosmological standard model ($\Lambda$CDM model) that is based on the Cosmological Principle,
the distribution of galaxies is quite homogeneous and isotropic when averaged over very large scales, with an intrinsic dipole in its number count determined by the density fluctuations, velocity and gravitational potential on large scales \ch{using linear perturbation theory}~\cite{Bonvin:2011bg,Challinor:2011bk}.
In Fig.~\ref{f:intrinsic}, we show the expected amplitude of the intrinsic dipole $d_\mathrm{int}=\sqrt{9C_1^\mathrm{int}/(4\pi)}$ (see e.g.~\cite{Gibelyou:2012}) for different tanh-smoothed top-hat redshift distributions with redshift limits $z_\mathrm{min}$ and $z_\mathrm{max}$. The intrinsic angular power spectrum $C_\ell^\mathrm{int}$ is calculated by the public Boltzmann code {\sc class}~\cite{Blas:2011rf} updated to {\sc class}gal~\cite{DiDio:2013bqa}
using the halofit power spectrum for the matter distribution~\cite{Takahashi:2012em}. In the {\sc class}gal code, we simply replaced $\ell_{\min}=2$ by $\ell_{\min}=1$ which does not include local terms like the kinematic dipole.
The other local terms are monopoles and therefore irrelevant for our discussion \ch{(see Appendix~\ref{app:Dip-int} for more details)}.

On very small scales, the dipole is expected to be large and the halofit model is not adequate especially for modelling velocities. Above $z\simeq 0.1$ where we average over more than 100Mpc, however, we expect this intrinsic dipole to be reasonably accurate. From Fig.~\ref{f:intrinsic} we see that the intrinsic dipole typically is of order $10^{-3}$.
It is important to note that up to redshift about $z\simeq 1$ the intrinsic dipole is comparable to the kinematic one. Due to the scale invariant power spectrum, fluctuations decay only very slowly at large scales.

The final contribution to (\ref{eq:combined_dipole}) comes from the shot noise (Poisson noise) \cite{Crawford:2009_SN,Rubart:2013tx} which is due to the fact that we sample the density field with discrete sources, typically galaxies.
The angular power spectrum due to shot noise is
\be
C_\ell^\mathrm{SN} = 1/\ndensbar \, ,
\label{eq:C_SN}
\ee
where $\ndensbar =\Ntot/(4\pi f_{\rm sky})$ denotes the angular number density in the survey, $\Ntot$ is the total number of sources and $\fsky$ is the observed fraction of the sky.
Using this result, we find the typical shot noise dipole on a full sky ($f_{\rm sky}=1$) to be 
\be
d_\mathrm{SN}^N = \sqrt{\frac{9C_1^\mathrm{SN}}{4\pi} } = \frac{3}{\sqrt{\Ntot}} \,.
\label{eq:dSN}
\ee
For a survey observing $10^6$ sources, one expects a shot noise dipole of $3\times 10^{-3}$, which again is of the same order of magnitude as the kinematic dipole.

In practice, it is not possible to use a survey covering the full sky; instead, only a fraction $\fsky$ of the full sky is observed. 
\ch{If the data contains only a monopole and a dipole, our estimator which we define in Section~\ref{s:estimator} is still exact. However, if there are higher multipoles present, these can `leak into' the estimated dipole.}
For the intrinsic dipole and the shot noise dipole, there will be a leakage of power from higher multipoles into the dipole.
\ch{If $\fsky < 1$, the amplitudes of the two contributions $\dint$ and $\dsnNC$ are in general larger than $\sqrt{9 C_1^\mathrm{int} / 4\pi}$ and $\sqrt{9 C_1^\mathrm{SN} / 4\pi}$, respectively}. A reasonable approximation of this effect is to replace $C_1^\mathrm{int/SN}$ by $C_1^\mathrm{int/SN}/\fsky$. For the intrinsic dipole, this assumes that the angular power spectrum $C_l^\mathrm{int}$ of the intrinsic anisotropies is relatively smooth, which is the case for the clustering expected in $\La$CDM.
Different survey geometries also lead to biased directions of the intrinsic and shot noise dipole contributions \ch{because of the leakage effect}.
\ch{We use simulated mock catalogs with the proper sky coverage of the surveys that fully include both the above consequences, the increase of amplitude and the directional bias, in the results section of this paper}.
\ch{Note that the effect of a partial sky coverage on the kinematic dipole is negligible since all higher multipoles, which could leak into the dipole due to the presence of a mask, are at least a factor of $\beta$ smaller than the kinematic dipole itself.}

\subsection{Statistics}
\label{s:statistics}

A measurement of the dipole in source number counts contains information about the sum of the three vectors in \eqref{eq:combined_dipole}.
In the following we address the question what to expect for the amplitude and orientation of the combined dipole $\dipole^N$, given its three individual contributions $\dipole_{\rm kin}, \dipole_{\rm int}$ and $\dipole_{\rm SN}$
(in this section we drop the superscript $N$ for number count, so for example $\dipole=\dipole^N$).
\ch{
The results derived here hold exactly for the full sky. 
They are nevertheless useful for practical applications with partial sky coverages for two reasons. First, they let us understand the impact of a random dipole on the dipole we aim to measure, i.e., we understand the influence of shot noise and of the intrinsic anisotropy on the accuracy and precision with which we can measure the kinematic dipole.
Secondly, the results obtained here are used later to extrapolate our simulation results (which are on the partial sky) to a larger number of sources. Doing so, we introduce one free parameter to account for the effects of the partial sky coverage.
We will come back to this at the end of \secref{s:new_SN}.
}

Both the intrinsic as well as the shot noise dipole point in independent random directions, with their amplitudes determined by the respective dipole powers $C_1^\mathrm{int}$ and $C_1^\mathrm{SN}$.
\ch{In a standard $\La$CDM cosmology with Gaussian initial conditions,  nonlinear clustering plays a negligible role for the dipole. Therefore,} the spherical harmonic coefficients with $\ell=1$, and hence the intrinsic dipole components are well described as Gaussian variables.
For a large number of sources, by the central limit theorem, the shot noise dipole can also be described as Gaussian.
Therefore, each of the components of the two dipoles $\dipole_{\rm int}$ and $\dipole_{\rm SN}$, (given by $d^{x}_\mathrm{int/SN}, d^{y}_\mathrm{int/SN}$ and $d^{z}_\mathrm{int/SN}$) are independent Gaussian random variables with zero mean and standard deviation $\sigma_\mathrm{int/SN}$ defined as
\begin{equation}
\sigma_\mathrm{int/SN} = \sqrt{\frac{3 C_1^\mathrm{int/SN}}{4\pi}}
\, ,
\label{eq:sigma_SN}
\end{equation}
such that 
\begin{equation}
\langle (d_\mathrm{int/SN})^2 \rangle
=
\langle (d^{x}_\mathrm{int/SN})^2 +
(d^{y}_\mathrm{int/SN})^2 +
(d^{z}_\mathrm{int/SN})^2 \rangle
= \frac{9 C_1^\mathrm{int/SN}}{4\pi} \, .
\end{equation}
In this section, we define the $z$-direction as the direction of the kinematic dipole. The components $d^{x/y/z}$ of the combined dipole $\dipole$ from \eqref{eq:combined_dipole} also follow a Gaussian distribution with variance
\begin{equation}
\sigma_r^2 = \sigma_\mathrm{int}^2 + \sigma_\mathrm{SN}^2
= \frac{3}{4\pi} (C_1^\mathrm{int} + C_1^\mathrm{SN})
\end{equation}
and means 
$\langle d^{x/y} \rangle = 0$, $\langle d^{z} \rangle = d_\mathrm{kin}$.
The amplitude $d$ thus follows a non-central $\chi(3)$-distribution, whose mean and variance
\ch{are computed in Appendix~\ref{app:derivation_non_central} using Refs.~\cite{Nuttall_QM,noncentral_chi2,confluent_hypergeometric}. They can be expressed with the associated Laguerre polynomial $L^m_n$ as}
\begin{equation}
\begin{aligned}
\langle{d}\rangle &=
\sigma_r \sqrt{\frac{\pi}{2}} L_{1/2}^{1/2} \Big(- \frac{d_\mathrm{kin}^2}{2 \sigma_r^2}\Big) \, ,
\\
\sigma_d^2 &=
3 \sigma_r^2 + d_\mathrm{kin}^2 - \langle{d}\rangle^2
\, .
\end{aligned}
\label{eq:mean_and_sigma_d}
\end{equation}

The direction of the combined dipole is completely described by the angle $\theta$ between the kinematic dipole and the combined dipole, which we call the deviation angle of the number count dipole.
To compute its mean and standard deviation, we consider the intrinsic dipole and the shot noise dipole as a single random dipole,
with  amplitude $d_r$ and the direction characterized by the angle $\theta_r$ between the kinematic dipole and the random dipole.
The random dipole amplitude is distributed accordingly to a (central) $\chi(3)$ distribution
\be
P(d_r,\sigma_r)=\frac{d_r^2}{\sigma_r^3}
\exp \Big( 
\frac{- d_r^2}{2 \sigma_r^2}
\Big)
\sqrt{\frac{2}{\pi}} \, .
\ee
Since we assume a full sky coverage, the random dipole's direction is  distributed uniformly,  $P(\cos \theta_r) =1/2$. The resulting deviation angle is given by
\begin{equation}
\theta =
\arccos (\uvec{d}_\mathrm{kin} \uvec{d})
=
\arccos \Big(\frac{d_r \cdot \cos \theta_r + d_\mathrm{kin}}{\sqrt{d_r^2 + d_\mathrm{kin}^2 +2 d_r d_\mathrm{kin} \cos{\theta_r}}}\Big) \ ,
\end{equation}
so that we can numerically compute its mean and variance
\begin{equation}
\begin{aligned}
\langle{\theta}\rangle &=
\frac{1}{2}\int_{-1}^{1} \mathrm{d}\cos{\theta_r} 
\int_{0}^{\infty} \mathrm{d} d_r
P(d_r,\sigma_r) \, 
\theta
\, , \\
\sigma_\theta^2 &=
\frac{1}{2}\int_{-1}^{1} \mathrm{d}\cos{\theta_r} 
\int_{0}^{\infty} \mathrm{d} d_r
P(d_r,\sigma_r) \, 
(\theta-\langle \theta \rangle)^2 \, .
\label{eq:mean_sigma_theta}
\end{aligned}
\end{equation}
This formalism leading to \eqref{eq:mean_and_sigma_d} and \eqref{eq:mean_sigma_theta} can be extended for a general signal dipole instead of $\dkin$ and a general source of noise $\sigma_r$, a point to which we return in \secref{s:new_SN}.

\begin{figure}
\begin{center}
\includegraphics[width=\textwidth]{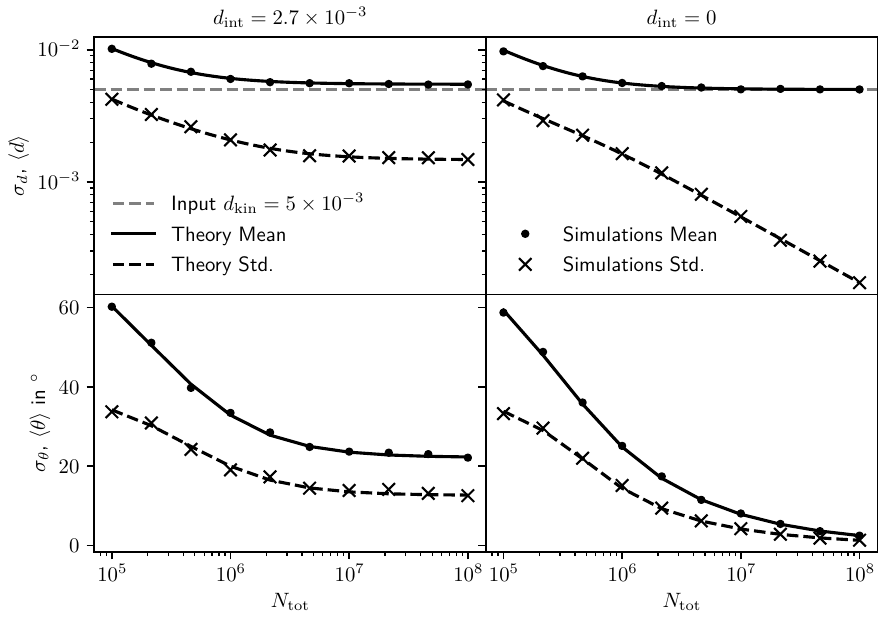}
\end{center}
\caption{
Means (dots and solid black lines) and standard deviations (crosses and dashed black lines) of the dipole amplitude (top panel) and deviation angle  in degrees (lower panel) as a function of the total number of sources, $N_{\rm tot}$. Lines are calculated following \eqref{eq:mean_and_sigma_d} and \eqref{eq:mean_sigma_theta}, and dots and crosses by using the estimator introduced in Section \ref{s:estimator} to evaluate 1000 realizations of a sky including the intrinsic dipole, $d_\mathrm{int}=2.7\times 10^{-3}$ ({left}) and $d_\mathrm{int}=0$ ({right}), shot noise depending on $\Ntot$ and the kinematic dipole ($d_\mathrm{kin}=5\times 10^{-3}$).
See main text for discussion.
}
\label{f:statistics}
\end{figure}

The means and standard deviations of the dipole amplitude and the deviation angle calculated from \eqref{eq:mean_and_sigma_d} and \eqref{eq:mean_sigma_theta},  are shown as the solid and dashed lines in \figref{f:statistics} as a function of the total number of sources, $N_{\rm tot}$. The dots and crosses in this figure are obtained from simulations using the estimator discussed in the next section.
The level of agreement indicates the validity of the analytical results to correctly describe how the shot noise and intrinsic dipole influence the estimated dipole.
It is worth noting that the amplitude of the combined dipole is biased towards larger values than the kinematic dipole amplitude based on the amplitude of the random dipoles.

The left panels of \figref{f:statistics} show the results for an intrinsic dipole of $d_{\rm int} = 2.7 \times 10^{-3}$, where we observe that even for $N_\mathrm{tot} \ra \infty$, neither the mean deviation angle nor the standard deviation of the dipole amplitude converge to zero, but instead to a finite value depending on the intrinsic dipole.
This means that intrinsic clustering imposes a systematic limit in the accuracy of observing the direction of the kinematic dipole as well as in the precision with which the kinematic dipole amplitude and thus our velocity can be determined.
The importance of the intrinsic dipole has been discussed in previous studies (e.g.~of the NVSS~\cite{Tiwari:2016_Cluster_Simulation,Bengaly:2017slg}) and the SKA forecast~\cite{Bengaly_2019}, where the intrinsic dipole is found to be the limiting factor for observing the kinematic dipole amplitude.
\subsection{Estimator}
\label{s:estimator}
\newcommand{\mon}{m}
\newcommand{\map}{N}
We use the \fitdipole method in \healpy and \healpix \footnote{\url{http://healpix.sf.net/} and \url{https://healpy.readthedocs.io/en/latest/\#}} \cite{healpy1,healpy2} as our dipole estimator.
\healpix provides a pixelisation scheme of the sphere, where each pixel covers the same area. The estimator minimizes the quantity $\sum_p (\map_p-\myvec{D} \cdot \myvec{r}_p)^2$ to find the best fit variables for the monopole and the three dipole components parameterized in the 4-component vector $\myvec{D} =(\mon,D_x,D_y,D_z)$, where the scalar product is taken with $\myvec{r}_p = (1 ,\uvec{r}_p)$ and $\uvec{r}_p$ denotes the unit vector in direction of the pixel $p$.
The sum runs over all unmasked pixels and $\map_p$ is the number of sources within the flux cut observed in the pixel $p$.
The solution for the best fit monopole and dipole is the linear equation
\begin{align}
\myvec{D}
=
\myvec{M}^{-1}
\sum_{p }
\myvec{r}_{p} \map_p
\, , \qquad
M_{ij} =
\sum_{ p }
r_{p,i} r_{p,j}
\, .
\label{eq:M_definition}
\end{align}
Since we define the dipole to characterize the relative anisotropy \eqref{eq:dipole_definition}, we need to divide by the monopole.
Our estimator for the number count dipole is 
\be
\estNC = (D_x/m, D_y/m, D_z/m)
\, .
\label{eq:estNC}
\ee
This estimator is used to compare the theory with simulations in \figref{f:statistics}.

\ch{We stress that, if the underlying field only contains a monopole and a dipole, this estimator is exact and not affected by the presence of a mask in the data. However, if higher order multipoles, such as a quadrupole, octopole etc., are present, these can look like a dipole on a partial sky coverage, and it is possible that the dipole measure differs from what would be observed on the full sky. This is the case for the intrinsic dipole and the shot noise dipole, but not for the kinematic dipole where higher order multipoles are negligible (see discussion at the end of \secref{s:EB_derivation}).}

The quadratic estimator advocated in \cite{siewert:2020_quad_est} only differs by
the normalization of the summand, minimizing
$\chi^2 = \sum_p (\map_p-\myvec{D}\cdot \myvec{r}_p)^2/(\myvec{D}\cdot \myvec{r}_p)$.
This is more appropriate to use if the number of sources per pixel differ significantly from pixel to pixel, because the normalization properly takes the shot noise into account.
Since the number of sources per pixel does not vary by a large amount in our present case, we use the linear estimator in what follows.\footnote{We tested the two estimators for both complete and incomplete sky coverages.
On simulated shot noise fields, the dipole amplitudes and directions found by the two estimators agree for each realization of shot noise within on average $6\%$ and $5\degree$, respectively.
Also, the mean and variance of all observed outcomes are equal for both estimators.
Therefore, for the relevant applications, the two estimators are unbiased with respect to each other and perform equally well.}

\section{A new method}\label{s:new}

We now introduce a new method to separate the kinematic dipole from the intrinsic dipole. In an inhomogeneous cosmological model the latter can be part of the background cosmology, but in a perturbed Friedman-Lema\^\i tre universe -- the case we are most interested in -- the intrinsic dipole is equal to the large scale clustering dipole.
Apart from allowing a measurement of our local velocity independent of any kind of number count anisotropy, our new method is also able to measure the anisotropy itself to test whether there is a significant intrinsic dipole due to, e.g.,  spacetime not being isotropic as postulated in~\cite{Secrest_2021}.

The idea to separate these two contributions is also explored in \cite{Tiwari:2015_NVSS}. In this reference, the dipole in sky brightness of the NVSS sources was computed, i.e., the dipole in the distribution of sources weighted by their flux.
Both, this dipole and the number count dipole contain the intrinsic dipole, but the kinematic contributions enter differently in these two dipoles, which we discuss below.
Therefore, measuring the number count dipole and the sky brightness dipole together can be used to separate the kinematic contribution to the dipole from the intrinsic dipole.
This allows to measure our velocity and the intrinsic dipole independently.
Since the cumulative source count above a flux limit follows a power law very closely, the kinematic dipole in sky brightness is very similar to the kinematic dipole in the number counts~\cite{Tiwari:2015_NVSS}.
This makes the uncertainty in estimating $\vecbeta$ and $\dint$ fairly large (see \secref{s:new_SN}).

Here, we present an extended and more general approach by using the fact that quantities other than the fluxes and the positions of the sources are changed due to our local motion
(leading to the number count dipole). It is obvious that the redshift $(1+z)$ is altered because of the Doppler effect by a factor $\delta(\uvec{n})^{-1}$, with $\delta(\uvec{n}) = (1+\uvec{n} \cdot \vecbeta)/\sqrt{1-\beta^2}$.
Spectroscopic and photometric redshift measurements thus contain a direction-dependent bias because of our local motion, which can be used to find kinematic signals arising from our motion in the survey data\footnote{We note that this effect has also been employed in a related approach~\cite{Singal_2019}.}.
Similarly to the redshift, the angular size $A$ of each source changes due to the special relativistic aberration in \eqref{eq:solid_angle_change},
so $A = \delta(\nvec)^{-2} {A}_\mathrm{rest}$.
Also, the observed one-dimensional angular size of a source, which we call $\phi$, is different from that in the rest frame by $\phi = \delta(\nvec)^{-1} \phi_\mathrm{rest}$.
This is true for any orientation of the source but only for very small sizes $\phi \ll 1$. This condition is clearly fulfilled since the sizes of galaxies are of order arcseconds.
Observationally, $\phi$ can be the fitted major axis of a source.

We summarize how flux, redshift and size are affected by our peculiar velocity depending on the angle
between our direction of motion and the source position
$\cos \theta =\uvec{\beta}\cdot\uvec{n}$:
\begin{gather}
\{F,1+z,\phi\} = \delta_{F,z,\phi}  \rframe{\{F,1+z,\phi\}} \\
\begin{aligned}
\delta_F &= 
\delta(\nvec)^{1+\alpha} \approx 1+(1+\alpha)\beta \cos \theta \\
\delta_{z} &= \delta(\nvec)^{-1} \approx 1-\beta \cos \theta \\
\delta_\phi &= \delta(\nvec)^{-1} \approx 1-\beta \cos \theta 
\, .
\label{eq:deltas}
\end{aligned}
\end{gather}
While there may be other pertinent observables, we concentrate on the above three in the present work.
The change of flux is equivalently expressed as a change in the apparent magnitude $m=-2.5 \log (F/F_0)$ defined with respect to a reference flux $F_0$ as
\begin{equation}
    m_\mathrm{} = m_\mathrm{rest} - 2.5 \log \delta_F(\uvec{n})
    \approx
    m_\mathrm{rest}-\frac{2.5}{\ln 10}
    (1+\alpha) \beta \cos \theta
    \, .
    \label{eq:magnitude_delta}
\end{equation}

\subsection{Weighted dipoles}
\label{s:weighted_dipoles}

Let us consider the distribution of the sources weighted by some function, $W$,  which can depend on any combination of properties that are measured for each source.
Thereby, we extend the number density \ch{$\ndens(\nvec)$} to the weighted field
\begin{align}
    \ndensW(\nvec)
    & \defined \int_\mathrm{cut}
    \mathrm{d}\ndens (\uvec{n}, F,z,\phi, ... ) W(F,z,\phi, ...) 
    \label{eq:Nw_int}
    \\ 
    & = \sum_{i \in \mathrm{cut}} W_i(F_i,z_i,\phi_i, ...) \delta(\uvec{n}_i - \uvec{n})
    \label{eq:Nw}
    \\ 
    & \defined ( 1 + \dipoleW \cdot \uvec{n} ) \ndensWbar +\order{\hat{n}_j^2} \, .
    \label{eq:Nw_dw}
\end{align}
Here, we denote the observed number density of sources in a direction $\nvec$ per flux, redshift, and angular size (and possibly other properties) as $\mathrm{d}\ndens (\uvec{n}, F,z,\phi, ... )$.
The integral in \eqref{eq:Nw_int} implicitly includes all the cuts in any arbitrary combination of the observed properties, i.e.~one integrates over the range of properties within all possible thresholds, which include a minimum flux cut to express the telescope sensitivity.
Introducing the Dirac delta function $\delta(\nvec_i-\nvec)$, the sum in \eqref{eq:Nw} runs over all sources $i$ within the cuts.
From this sum, by averaging over small angular regions, the field $\ndensW (\uvec{n})$ can be reconstructed from observations.
Of all the anisotropies on different scales in this weighted field, only the weighted dipole $\dipoleW$ is of interest to us and higher multipoles $\order{\hat{n}_j^2}$ are ignored.
Similarly to \eqref{eq:dipole_definition}, we define the weighted dipole relative to the mean value of the field, $\ndensWbar$,
\ch{which generally differs from the mean number density, $\ndensWbar \neq \ndensbar$.}

For $W=1$, the weighted field reduces to the number density and the weighted dipole to the number count dipole derived in \secref{s:EB_derivation}, $\dipole^{W=1}=\dipoleNC$.
The case $W=F$ corresponds to the dipole in sky brightness, as computed in \cite{Singal:2011} and \cite{Tiwari:2015_NVSS}.
Both the number count dipole and any weighted dipoles contain a kinematic contribution, the intrinsic clustering dipole and a random shot noise dipole
\begin{align}
    \dipoleNC   &= \dkinNC + \dint + \dsnNC 
    \label{eq:dNC_and_dW_1} \\ 
    \dipoleW &= \dkinW + \dint + \dsnW \,. 
    \label{eq:dNC_and_dW_2}
\end{align}
Although both dipoles contain the same sources, the respective contributions from shot noise, $\dsnNC$ and $\dsnW$, are different because of the random distribution of the weights (see \secref{s:new_SN}).

Of the three contributions, the intrinsic dipole $\dint$ is the only one that is equal in both expressions \eqref{eq:dNC_and_dW_1} and \eqref{eq:dNC_and_dW_2}, because both include exactly the same sources and their clustering.
\ch{Here, we assume that the properties of a source are independent of the source's position. While this is true for the properties in the rest frame of each source, it is not exact for the observed properties of the sources.
These can be intrinsically biased in different directions on the sky, e.g. through velocity or gravitational potentials.}
\ch{
For one, this leads to fluctuations in the number density which are included in our simulations (see Appendix~\ref{app:Dip-int}). Additionally, the weights that depend on the flux, size and redshift might differ somewhat in different directions even for an observer at rest, which is not accounted for.
In the following, we argue by which amount this simplification can systematically affect the measurement.
To do so, we exemplarily investigate the bulk flow velocity, defined as the root mean square velocity of all sources with respect to the background cosmology.
}
\ch{
The bulk flow changes the flux, redshift and size due to the Doppler effect and therefore induces an intrinsic dipole in the weights.}
We calculate the bulk flow in Appendix~\ref{app:bulk_flow} for the applications we consider here, finding a velocity of approximately $2\times 10^{-5}$,
less than $2\%$ of our expected $\beta$.
Other effects \ch{that can also lead to a dipolar bias of the weights}, e.g., from gravitational potentials or lensing are expected to be of similar order or even smaller.
Therefore, in the $\Lambda$CDM model, neglecting \ch{the intrinsic dipole in the}  weights leads to a systematic uncertainty of approximately $2\%$ in comparing our velocity with the velocity inferred from the CMB dipole.

The kinematic contributions of both the number count dipole and the weighted dipole, $\dkinNC$ and $\dkinW$, point in the same direction as $\vecbeta$.
Thus, we only need to find their amplitudes $\ampNC$ and $\ampW$ defined via $\dkinW = \ampW \vecbeta$, $\dkinNC = \ampNC \vecbeta$.
By assuming $N(>\Flim) \propto \Flim^{-x}$, we found that $\ampNC = 2+(1+\alpha)x$ in \secref{s:con}; however, it is not possible to find an analytic expression for $\ampW$ without making many more assumptions about the dependence of $\mathrm{d} \ndensbar(F,z,\phi,...)$ on the other quantities.
Instead, we find the amplitudes numerically using the catalog of observed sources.
Even though these sources already contain the effects of our motion and some measurement error, they are assumed to be close to the true distribution\footnote{
Note that this is also a necessary assumption in the usual number count dipole analysis when fitting $N(>\Flim) \propto \Flim^{-x}$ in order to the data to find $x$.
}.
In \secref{s:applications}, we will see how observational errors lead to a bias in the computed amplitudes and therefore also in the estimated $\vecbeta$ and $\dint$.

\ch{
We propose the following method to compute $\ampW$ (and $\ampNC$ with $W=1$) from a simulated mock catalog (and also to determine it from the observed data when our approach is applied).
We pretend that all of the simulated (or observed) sources are situated in the forwards ($+$) or backwards ($-$) direction.
For these two cases, the properties of all the sources are boosted with an artificial velocity $\beta_\mathrm{test}$ accordingly to \eqref{eq:deltas} to find
}
\begin{equation}
\ndens^{\pm}_W=\sum_{i \in \mathrm{cut}} W_i(\delta_F^\pm F_i, \delta_z^\pm z_i, \delta_\phi^\pm \phi_i,...) \, ,
\label{eq:N_pm}
\end{equation}
where \ch{$\delta_{F,z,\phi}^\pm = \delta_{F,z,\phi}(\uvec{n} \cdot \vecbeta=\pm \beta_\mathrm{test})$}.
The implicit cut is imposed after boosting the sources in the two different directions, respectively.
$\ndens^{\pm}_W$ are proportional to the expected values of the weighted field in the forward ($+$) and backward ($-$) directions,
\ch{
i.e., they are proportional to the maximum and minimum value of the dipole field, respectively.
Therefore, we can find the expected amplitude 
\begin{equation}
    \ampW = 2 + \frac{\ndens^+_W - \ndens^-_W}{\ndens^+_W + \ndens^-_W} \beta_\mathrm{test}^{-1}
    \, .
    \label{eq:compute_bw}
\end{equation}
We add $2$ to accommodate the well-known change of the solid angle (see~\eqref{eq:solid_angle_change}) which is not included in this procedure.
}

\ch{
This way to obtain the amplitude does not depend on a mask, or any directional choice of the test velocity.
Only the test speed, $\beta_\mathrm{test}$, needs to be fixed.
While there are some random fluctuations in the computed amplitude when using different values of $\beta_\mathrm{test}$, these are tested to be smaller than $0.5\%$ for $10^{-3}<\beta_\mathrm{test}<10^{-2}$.
In principle, one can also compute the amplitudes for a prior distribution of $\beta_\mathrm{test}$ amplitudes and marginalize over it to obtain a best estimate.
For simplicity, in this work we fix $\beta_\mathrm{test}=0.002$.
We have found that this empirical method to find $B_W$ is very robust. Another possibility would of course be to determine the derivative of $\ndensWbar$ w.r.t. the properties (flux, redshift, angular size) considered. The method proposed here is actually just a simple procedure to do this in practice.
}

We now show how by subtracting and adding the two observed dipoles \eqref{eq:dNC_and_dW_1} and \eqref{eq:dNC_and_dW_2}, it is possible to measure our velocity and the direction of our motion, as well as the amplitude and direction of the intrinsic clustering dipole separately.

\subsection{Estimator}
\label{s:new_estimator}
In order to measure the weighted dipole $\dipole_W$ in the field $\ndensW(\uvec{n})$, we generalize the estimator introduced in \secref{s:estimator}, Eq.~\eqref{eq:estNC}, by weighting every observed source $i$ with the quantity $W_i(F_i,z_i,\phi_i,...)$, which can depend on different properties of the specific source.
The estimator  $\myvec{d}^W_{\rm est}$ for the weighted dipole $\dipole^W$ in \eqref{eq:dNC_and_dW_2} is obtained by replacing the number of sources per pixel $N_p$ in \eqref{eq:M_definition} with $w_p = \sum_{i \in p} W_i$, i.e.,~the sum of the weights of all the sources in the pixel $p$. Again, the sum implicitly runs over all the sources that are above the minimum flux cut (and possible limits in other quantities).

Defining $\Delta = \ampW - \ampNC$, we introduce the estimators for our velocity and the intrinsic dipole
\begin{align}
\estBeta &= \frac{\estW - \estNC}{\Delta}
\label{eq:estbeta}
\\
\estInt &= \frac{\ampW \estNC  - \ampNC \estW}{\Delta} \, .
\label{eq:estdint}
\end{align}
Comparing these with equations \eqref{eq:dNC_and_dW_1} and \eqref{eq:dNC_and_dW_2}, we see that each of the above estimators eliminates the corresponding other contribution as desired. However, the shot noise is not removed; we calculate the remaining noise below.

\subsection{Shot noise}
\label{s:new_SN}

As can be seen by comparing \eqref{eq:Nw_dw} with \eqref{eq:estbeta}, to estimate $\vecbeta$, we are actually looking for the dipole in the `velocity field':
\begin{equation}
    \beta(\uvec{n}) \defined
    \frac{1}{\ndensbar \Delta}
    \left( 
    \overbar{W}^{-1}
    \ndensW(\nvec) -
    \ndens(\nvec)
    \right)
    = \frac{1}{\ndensbar \Delta} \sum_i \delta(\uvec{n}_i - \uvec{n})
    \left( \frac{W_i}{\overbar{W}} -1\right)
\end{equation}
Here we have introduced the mean weight $\overbar{W} = \ndensWbar/\ndensbar$.
The dipole in $\beta(\nvec)$ is $\vecbeta$ plus a shot noise contribution, which is found by calculating the constant angular power spectrum of $\beta(\nvec)$:
\footnote{
In general, for a field
$f(\nvec)=\sum_i \delta(\uvec{n}_i - \uvec{n}) f_i /\ndensbar$,
where any two of all $f_i$ and $\nvec_j$  are uncorrelated,
the angular power spectrum for $\ell\geq 1$ is
$C_\ell=\frac{1}{2\ell+1}\sum_m \sum_{i,j}
\expect{Y_{\ell m}(\nvec_i) Y^\ast_{\ell m}(\nvec_j) f_i f_j}/\ndensbar^2
=\expect{f^2}/\ndensbar$,
which gives the results \eqref{eq:C_SN}, \eqref{eq:C_beta} and \eqref{eq:C_int}
for $f_i=1$, $(W_i/\overbar{W}-1)/\Delta$ and $(\ampW-\ampNC W_i/\overbar{W})/\Delta$, respectively.
}
\begin{equation}
    \CbetaSN= 
    \frac{1}{\ndensbar}
    \left( \frac{\sigma_W}{\overbar{W} \Delta} \right)^2 \, .
    \label{eq:C_beta}
\end{equation}
We defined the variance of all the observed weights $\sigma_W^2\defined \expect{(W-\overbar{W})^2}$.
Using \eqref{eq:mean_and_sigma_d} and inserting $\sigma_r^2 = 3\CbetaSN/4\pi$ gives the expected value and the variance of our velocity estimator \eqref{eq:estbeta}
\begin{equation}
\begin{aligned}
    \expect{\beta^\mathrm{est}} &= \sqrt{\frac{3\CbetaSN}{8}} L_{1/2}^{1/2} \Big(- \frac{2\pi \beta^2}{3\CbetaSN}\Big) 
    \\
    \sigma^2(\beta^\mathrm{est}) &= \frac{9}{4\pi} \CbetaSN + \beta^2 - \expect{\beta^\mathrm{est}}^2
    \, .
\end{aligned}
\label{eq:mean_and_sigma_beta}
\end{equation}
Here, $\beta$ is the true velocity, which in this paper (as well as in the $\Lambda$CDM model) is assumed to be the CMB velocity. From \eqref{eq:C_beta}, we see that the shot noise in $\beta$ is different from that in the number counts dipole in \eqref{eq:C_SN} by a factor $\sigma_W/\overbar{W}\Delta$.
The additional $\Delta$ appears because we already converted the dipole amplitude to a velocity.
Therefore, we are looking for a weight function $W(F,z,\phi, ...)$, whose values when evaluated for all sources have a small spread (small $\sigma_W/\overbar{W}$) and at the same time gives a large difference $\Delta$ between the two dipole amplitudes.
We define the norm of this quantity, which we want to choose to be as large as possible, by:
\begin{equation}
    \dsigW \defined
    \left| \Delta \frac{\overbar{W}}{\sigma_W} \right|  \, ,
\end{equation}
so the shot noise in the velocity field is simply $\CbetaSN = 1/(\ndensbar \dsigW^2)$.
The typical shot noise contribution to $\beta$ is $3/(\dsigW\sqrt{\Ntot})$, given a total number of $\Ntot$ sources.
The signal-to-noise ratio for such a measurement of $\beta$ is 
\begin{equation}
    \mathrm{S/N}= \dsigW  \beta  \sqrt{\Ntot}/3
    \, .
    \label{eq:signal_to_noise_beta}
\end{equation}
The signal-to-noise can be optimised either by modifying the weight function, or by adjusting the cuts that are applied to the data, e.g., the minimum and maximum flux.
Here, the trade-off is between cuts that constrain the range of possible weights leading to a small variance (large $\dsigW$) and the resulting smaller number of sources $\Ntot$.\footnote{By splitting the data in several subsets, finding an optimal weight function for each of these and averaging over the results, it might be possible to further improve the method. We do not test this possibility here.}

For the shot noise of the intrinsic dipole estimator, we find
\begin{equation}
    \CintSN= 
    \frac{1}{\ndensbar}\left[1+\left(\frac{ \ampNC}{\dsigW}\right)^2\right]
    \label{eq:C_int}
\end{equation}
and the mean and variance of the amplitude of the intrinsic dipole estimator \eqref{eq:estdint}
\begin{equation}
\begin{aligned}
    \expect{d^\mathrm{est}_\mathrm{int}} &= \sqrt{\frac{3\CintSN}{8}} L_{1/2}^{1/2} \Big(- \frac{2\pi d_\mathrm{int}^2}{3\CintSN}\Big) 
    \\
    \sigma^2(d_\mathrm{int}^\mathrm{est}) &= \frac{9}{4\pi} \CintSN + d_\mathrm{int}^2 - \expect{d^\mathrm{est}_\mathrm{int}}^2
    \, .
\end{aligned}
\label{eq:mean_and_sigma_int}
\end{equation}
Here, $d_\mathrm{int}$ is the true intrinsic dipole, which is assumed to be only the clustering dipole. To clarify, the quantity $\CintSN$ is not related at all to the intrinsic angular power spectrum.
Rather, it is the shot noise power spectrum of the field that we use to estimate $\dint$. The same arguments apply to $\CbetaSN$.

We are also able to calculate how much the estimated directions of the velocity and the intrinsic dipole deviate from the respective true values by using \eqref{eq:mean_sigma_theta}.
These are the deviation angles of the velocity $\theta_\beta$ and of the intrinsic dipole $\theta_\mathrm{int}$.
For this, we replace $d_\mathrm{kin}$ with $\beta$ and $\sigma_r^2$ with $3\CbetaSN/(4\pi)$ for the deviation angle of the velocity, and $d_\mathrm{kin}$ with the expected intrinsic dipole amplitude $d_\mathrm{int}$ and $\sigma_r^2$ with $3\CintSN/(4\pi)$ for the deviation angle of the intrinsic dipole.
With this, one obtains expressions for the means and standard deviations of our estimators for the deviation angles
\begin{equation}
    \expect{\theta_\beta} \, \, ,
    \qquad
    \expect{\theta_\mathrm{int}} \, \, ,
    \qquad
    \sigma(\theta_\beta) \quad \mathrm{and}
    \quad
    \sigma(\theta_\mathrm{int})
    \, ,
    \label{eq:mean_and_sigma_theta_int_beta}
\end{equation}
which we do not repeat here.

We have discussed the expected bias and precision of the estimators for our velocity and the intrinsic dipole regarding their amplitude and direction.
All the results in this section only hold for a full sky coverage, but also make clear how to optimize the estimator for any sky coverage. Additionally, they serve as a model that can be fit and extrapolated to a larger number of sources for the results from partial sky coverages \ch{(see also beginning of \secref{s:statistics})}.
We will come back to this when discussing the main results in \secref{s:results}.
\section{Applications}\label{s:applications}

In this section, we forecast the precision with which we can measure both, the local velocity $\vecbeta$ and the intrinsic dipole $\dint$ using the above weighting scheme in (i) an optical survey with redshift measurements and (ii) a survey measuring the size of radio galaxies.
We present the models for our mock catalogs, show how to find the optimal weights and  test the new approach in \secref{s:results}.

\begin{figure}
    \begin{minipage}{0.5\textwidth}
        \centering
        \includegraphics[width=\textwidth]{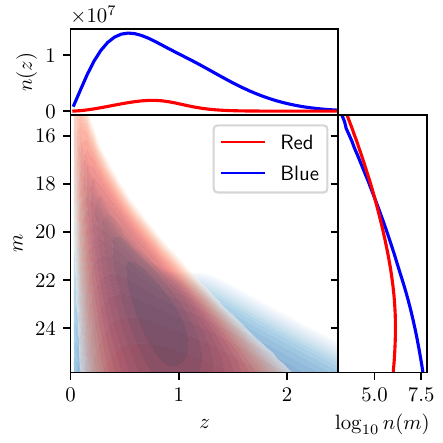}
    \end{minipage}\hfill
    \begin{minipage}{0.5\textwidth}
        \centering
        \includegraphics[width=\textwidth]{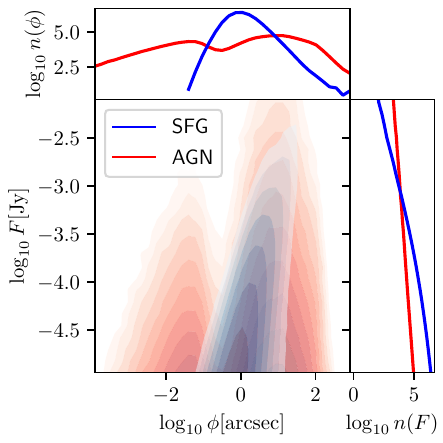}
    \end{minipage}
    \caption{\label{f:distributions} 
    Source distributions for the two surveys considered, LSST/Euclid (\textit{left}) and SKA (\textit{right}).
    The blue and red lines/contours indicate blue and red galaxies (\textit{left}) or star forming galaxies (SFG) and active galactic nuclei (AGN) (\textit{right}).
    The shaded contours show the source density as a function of redshift and magnitude [$n(z,m)$, left panel] and as a function of size and flux [$n(\phi,F)$, right panel] in arbitrary units on a logarithmic scale.
    A larger opacity corresponds to a larger source density.
    The side panels show the projected source densities (integrated over one of the two parameters).
    We can recognize the two different models for the size of the AGN depending on the viewing angle as described in~\cite{Bonaldi_2018} in the right panel.
    }
    \label{f:n_F_A_z}
\end{figure}

\subsection{LSST/Euclid including redshifts}
\label{s:redshift_survey}

We consider an optical galaxy survey similar to Euclid~\cite{Euclid_def_study} and Legacy Survey of Space and Time (LSST)~\cite{LSST_2009} planned with the Vera C.~Rubin Observatory, that observes $\Ntot$ galaxies measuring a r-band magnitude $m$ and redshift $z$.
We use the same procedure to create the mock catalogs for the LSST as well as for the Euclid survey,
following the LSST prescription.
The difference between the LSST and Euclid simulations is only in their sky coverage.
For Euclid, this corresponds to the photometric galaxy sample. As we will see later, having a large number of galaxies in the sample is much more important than a spectroscopic redshift precision, so we do not study the spectroscopic galaxy sample of Euclid here.

\paragraph{Source properties}
To create the mock catalogs we use the luminosity functions and redshift distributions modelled for the LSST as described in~\cite{Alonso:2015uua} for blue and red galaxies.
Their implementation which is used in this work is publicly available%
\footnote{
\url{http://intensitymapping.physics.ox.ac.uk/Codes/ULS/photometric/}
}.
We use approximately a Planck cosmology:
$$
\Om_m = 0.31 \qquad \Om_\La =1-\Om_m \qquad H_0 = 67.6\, {\rm km/s/Mpc} \,.
$$
We impose a magnitude limit of $m_\mathrm{lim}=26$~\cite{Alonso:2015uua}.
We start by computing the redshift distribution $n_\mathrm{b,r}(z)$ for blue and red galaxies, respectively, and normalize them to yield in total $\Ntot$ galaxies.
In $100$ redshift bins between $0$ and $3.5$, for each galaxy type, we draw $n_\mathrm{b,r}(z)$ random absolute magnitudes from the respective luminosity function at redshift $z$, under the condition that
they lead to an apparent magnitude larger than $m_\mathrm{lim}$ (using the luminosity distance and including $k$-correction, see~\cite{Alonso:2015uua}).
To each of these magnitudes, we associate a random redshift in the corresponding bin sampled from $n_\mathrm{b,r}(z)$.
In this way, we obtain a source distribution (see \figref{f:distributions}) that includes the correct correlation between redshift and apparent magnitudes.
In what follows, we no longer differentiate between red and blue galaxies.

\begin{figure}
    \centering
    \begin{minipage}{0.333\textwidth}
        \centering
        \includegraphics[width=\textwidth]{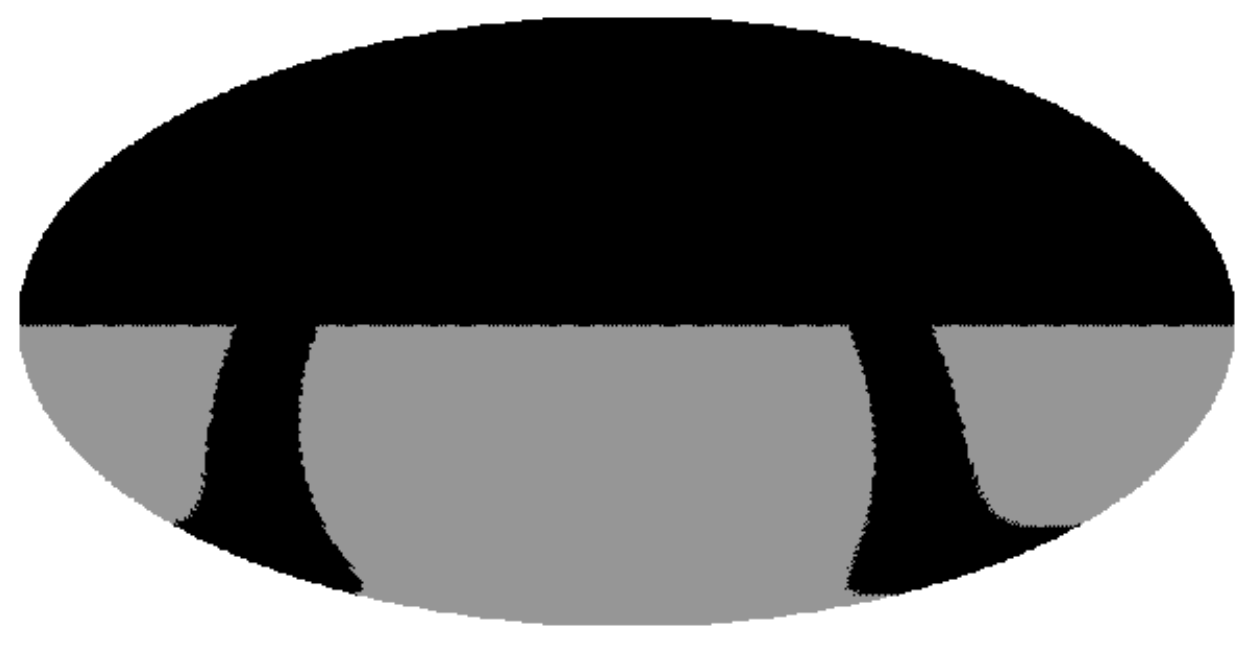}
    \end{minipage}\hfill
    \begin{minipage}{0.333\textwidth}
        \centering
        \includegraphics[width=\textwidth]{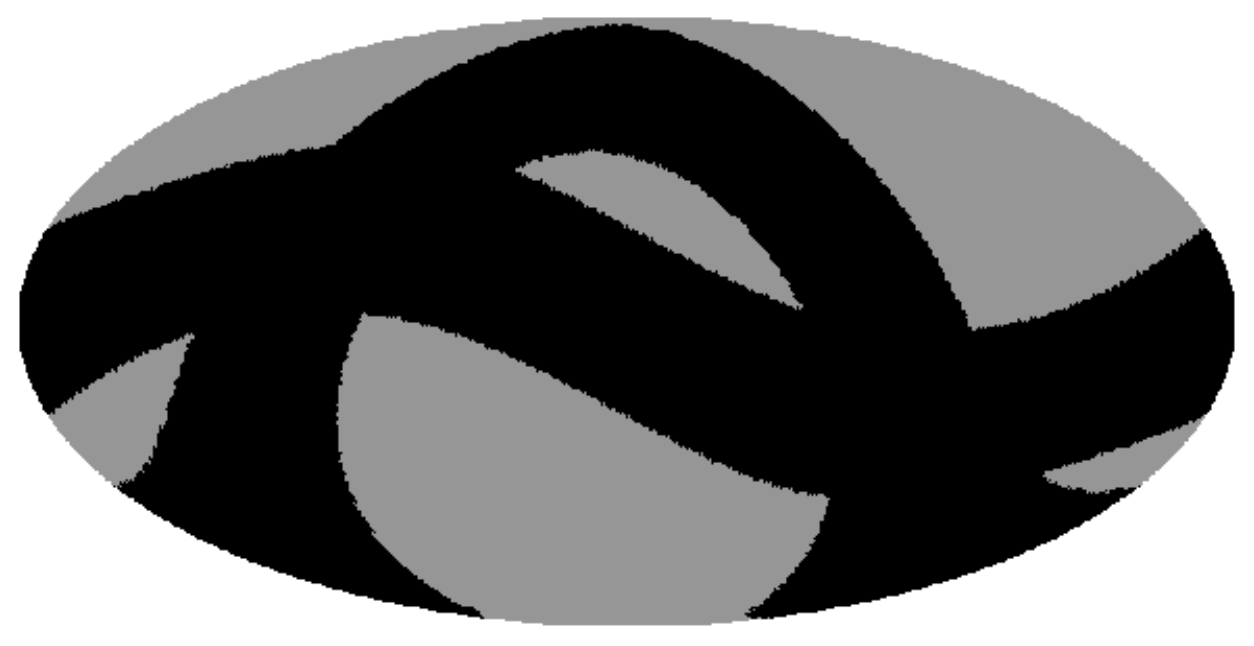}
    \end{minipage}%
    \hfill
    \begin{minipage}{0.333\textwidth}
        \centering
        \includegraphics[width=\textwidth]{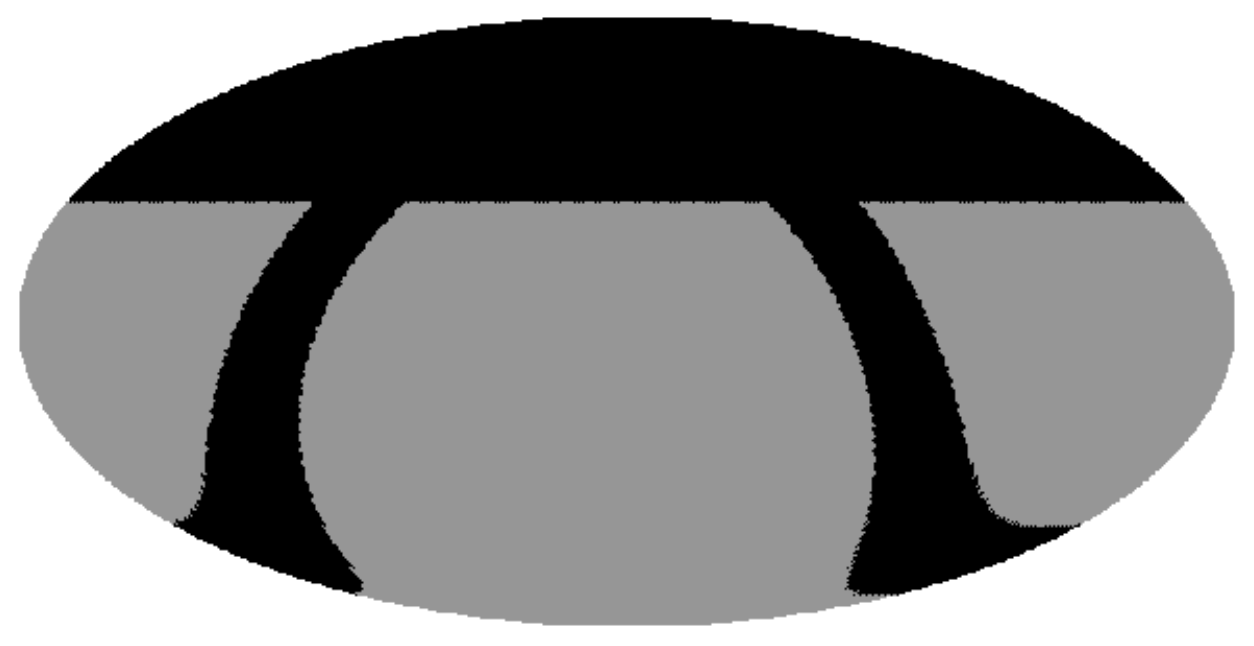}
    \end{minipage}%
    \caption{
    From left to right: LSST, Euclid and SKA masks in equatorial coordinates with sky coverages of
    $40\%$, $38\%$ and $61\%$,
    respectively.
    Grey indicates the observed regions and black the masked regions.
    The masks are described in the main text.
    }
    \label{f:masks}
\end{figure}

\paragraph{Mask} 
The LSST covers only the southern hemisphere in equatorial coordinates. We also mask the region within $\pm 10\degree$ of the galactic plane.
For Euclid, we mask the regions within $\pm 20\degree$ of the ecliptic plane and $\pm 20\degree$ of the galactic plane~\cite{Euclid_def_study}.
The masks are shown in \figref{f:masks}.

\paragraph{Realizing the Mocks}
To include the intrinsic large-scale clustering, we construct a Gaussian density field on the sphere by computing the angular power spectrum with {\sc class}gal~\cite{Blas:2011rf,DiDio:2013bqa} (including halofit~\cite{Takahashi:2012em}).
A Gaussian realization is a good approximation because non-linearities have only a small ($<1\%$) effect on large scales.
Sources are now distributed sampling this field. More precisely, we realize the density field on the sphere with the public code `\healpy'\footnote{\url{https://pypi.org/project/healpy/} and \url{http://healpix.sf.net/}}
in pixels of size $\sim 0.9\degree$ and then replace the field value with a random number of galaxies obeying Poisson statistics.

\begin{figure}
    \centering
    \includegraphics[width=\textwidth]{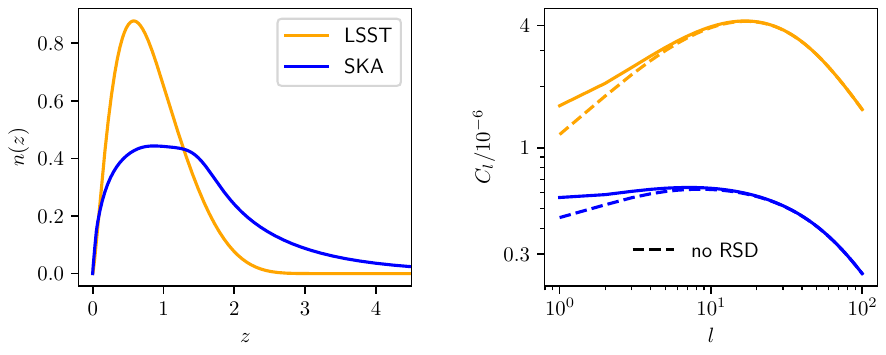}
    \vspace*{-5mm}
    \caption{
    \textit{Left:}
    The expected redshift distribution of LSST and SKA sources. 
    \newline
    \textit{Right:}
    The expected angular power spectrum for LSST and SKA (calculated with \class and following~\cite{Alonso:2015uua}).
    For the simulations, we only use $l<64$.
    Also shown is the angular power spectrum not including redshift space distortions (RSD) as dashed lines.
    This stresses their important contribution on large scales of nearly $30\%$.
    Note also that the dipole for SKA is about a factor 2 smaller than the one from LSST/Euclid.
    \label{fig:ang_power_spectrum_redshift}
    }
\end{figure}

The LSST angular power spectrum (\figref{fig:ang_power_spectrum_redshift},  right panel) is calculated with \class using the redshift distribution $n(z)=n_\mathrm{b}(z)+n_\mathrm{r}(z)$ (\figref{fig:ang_power_spectrum_redshift},  left panel) in the range $z=0$ to $z=3.5$, the galaxy bias $b(z) = 1 +0.84z$, the magnification bias and the evolution bias as given in~\cite{Alonso:2015uua}.
This angular power spectrum is also used for the Euclid simulations.
To this point, we have created a mock catalog in the rest frame of the sources.

Next, we include the kinematic effects due to our local motion by shifting each source position according to the special relativistic aberration (see Appendix A of \cite{Rubart:2013tx}) and boosting its redshift and magnitude.
To do so, we model the frequency dependence of the sources to be characterized by $F(m) \propto \nu^{-\OPTindex}$ (see discussion in \secref{s:rev}).
We assume that the distribution of $\OPTindex$ is known, for example from measurements of the magnitude in different frequency bands. Motivated by the colors shown, e.g., in \cite{colors_Korytov_2019},
we fix the spectral index to be 
$
    \OPTindex = 1
$.
With that, the kinematic effects due to our local motion are included by setting for each source
\begin{align}
    z+1 & ~\rightarrow~ (z+1) \delta(\nvec)^{-1} \\
    m & ~\rightarrow~ m - 2.5 \log \delta(\nvec)^{1+\OPTindex}
    \, .
\end{align}
Finally, redshift measurement errors are simulated by replacing the redshift for each source with the absolute value of a random normal variable $z \ra |\gaussian(z,\sigma_z)|$. The absolute value is used to avoid negative redshifts and cuts are applied afterwards.
For the main analysis, we impose a magnitude limit smaller than $m_\mathrm{lim}=26$ to avoid boundary effects and use all redshifts
\begin{equation}
    m<25.9 \,, \qquad  
    \mathrm{all} \, \, z 
    \, .
    \label{eq:cut_LSST}
\end{equation}

\subsection{SKA including source sizes}
\label{s:SKA_survey}

We consider the planned radio continuum survey by the SKA~\cite{SKA_redbook_2018}, in particular the measurements of the flux $F$ and the size $\phi$ of the observed sources.
We forecast for the SKA-2 survey, with a large sky coverage (see below) and a conservative minimum flux threshold of $F>10\mu$Jy~\cite{Bengaly_2019}.

\paragraph{Source properties}
As a model of the observed sources, we use the mock data from the \trecs simulations~\cite{Bonaldi_2018}.
These simulations model the flux and size of active galactic nuclei (AGN) and star forming galaxies (SFG).
We use all the AGN and SFG above $9\mu$Jy at $0.7$GHz in the `wide catalogue' of \trecs.
For the AGN, we use the projected apparent angular size of the core+jet emission as the size.
For the SFG, we use twice the projected apparent angular half light radius.
Both of these quantities correspond to the observable major diameter which we call $\phi$.
For each mock realization, we draw $\Ntot$ random tuples of flux and size from the combined data set of SFG and AGN (see right panel of \figref{f:distributions}).
In what follows, we do not distinguish between AGN and SFG.

\paragraph{Mask} 
We use the maximum possible sky coverage expected by the SKA-2 survey, masking the region above the equatorial declination of $30\degree$ and the region within $\pm 10\degree$ of the galactic plane~\cite{Bengaly_2019} (see right panel of \figref{f:masks}).

\paragraph{Realizing the mocks}
The angular power spectrum is calculated with \classgal and shown in \figref{fig:ang_power_spectrum_redshift}.
The redshift distribution, bias, magnification bias and evolution bias are all modelled by using the code\footnote{
\url{http://intensitymapping.physics.ox.ac.uk/codes.html}
}
developed in \cite{Alonso:2015uua}.
As in \secref{s:redshift_survey}, the angular power spectrum is used to distribute the sources on the sphere.
The kinematic effects \eqref{eq:deltas} are included by aberrating all positions and boosting the flux and size of each source:
\begin{align}
    F &\rightarrow F \delta(\nvec)^{1+\SKAindex}
    \label{eq:app_F}
    \\
    \phi &\rightarrow \phi \delta(\nvec)^{-1}
    \label{eq:app_A}
    \, 
\end{align}
with $\SKAindex=0.75$~\cite{SKA_redbook_2018}.
Unbiased measurement errors are included by replacing each size with a random normal variable $\phi\ra \gaussian(\phi,\sigma_\phi)$, and 
the cuts to size and flux are applied afterwards.
In this case, we choose
\begin{gather}
    0.3 \, \arcsec < \phi < 100\, \arcsec \, , \qquad
    10^{-5} \, \mathrm{Jy}< F <10^{-2} \, \mathrm{Jy}
    \, .
    \label{eq:cut_SKA}
\end{gather}
The minimum size and minimum flux are imposed because of the limiting resolution and sensitivity of the telescope.
Upper limits are chosen since they allow to find a better signal to noise ratio \eqref{eq:signal_to_noise_beta} by decreasing the variance in the weights.

\subsection{Optimal weights}

As can be seen from \eqref{eq:C_beta}, and \eqref{eq:C_int}, the shot noise in the estimation of both $\vecbeta$ and the intrinsic dipole $\dint$ is minimized when $\dsigW$ is maximized.
To find an optimal estimator, we therefore look for a weight function $W(z,m)$ (LSST/Euclid) and $W(\phi,F)$ (SKA) for which $\dsigW$ is at its maximum.
As explained previously, we use \eqref{eq:cut_LSST} and \eqref{eq:cut_SKA} for the cuts after finding reasonable values for $\Delta_W$ without discarding too many sources. We do not optimize on further cuts in the source properties.

For the LSST/Euclid simulations, we consider functions of the form $W(z,m) = (1+z)^{x_z} m^{x_m}$, weighting each source by a combination of its redshift and magnitude with the redshift exponent $x_z$ and the magnitude exponent $x_m$ as free parameters.
This gives good results and we leave the search for more complex weight functions that might result in larger values of $\dsigW$ to future work.

\begin{figure}
    \begin{minipage}{0.5\textwidth}
        \centering
        \includegraphics[width=\textwidth]{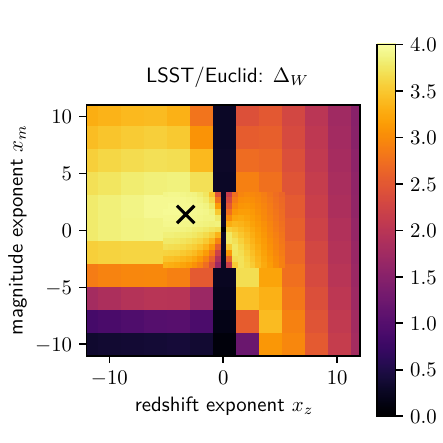}
    \end{minipage}\hfill
    \begin{minipage}{0.5\textwidth}
        \centering
        \includegraphics[width=\textwidth]{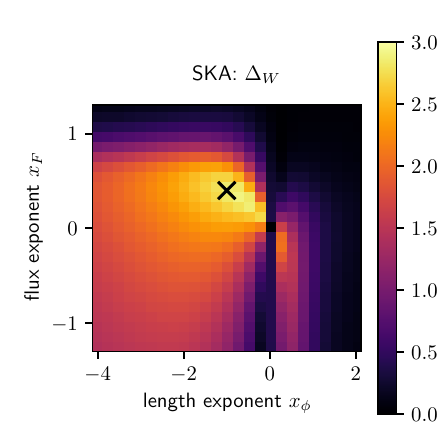}
    \end{minipage}
    \caption{\label{f:S2N_exponents}  
    We show$\dsigW=|\Delta\overbar{W}/\sigma_W|$,
    the ratio of the expected weight and its scatter (related to the signal to noise ratio by $\mathrm{S/N} = \dsigW \beta \sqrt{\Ntot}/3$), as a function of the weighting exponents.
    \newline
    \textit{Left:}
    LSST/Euclid: the weight function is $W(z,m)=(1+z)^{x_z} m^{x_m}$.
    The cross marks the maximum, which is found for $x_z=-3.3$ and $x_m=1.4$, where $\dsigW=3.91$. (Using all sources with $m<26$.)
    \newline
    \textit{Right:}
    SKA: the weight function is $W(\phi,F)=\phi^{x_\phi} F^{x_F}$. The cross marks the maximum, which is found for $x_\phi=-1$ and $x_F=0.4$, where $\dsigW=2.84$.
    (Using sources between $0.3<\phi[\text{arcsec}]<100$ and $-5<\log F[\text{Jy}] <-2$.)}
\end{figure}

In the left panel of \figref{f:S2N_exponents}, we show the value of $\dsigW$ for different exponents.
One can see that for $x_z=0$, i.e. weighting only by the magnitude, $\dsigW$ is very small.
This is related to the discussion from the beginning of this section:
a power law in the source count per flux interval $n(F)$ implies that the dipole weighted by any power of the flux is the same as the number count dipole~\cite{Tiwari:2015_NVSS}.
As can be seen in the left panel of \figref{f:distributions}, for our model, the number density per magnitude in fact closely follows a power law.
This is the reason why we also include the redshifts of the sources.
We find that there is a rather flat maximum at the optimal exponents,
\begin{equation}
    W(z,m) = (1+z)^{x_z} m^{x_m} \, , \quad x_z=-3.3, \quad x_m=1.4 \, , \quad \text{where   } \, \dsigW=3.91
    \, .
    \label{eq:exponents_LSST}
\end{equation}

We proceed analogously for the SKA mock catalogs.
A weight function of the form $W(\phi,F)=\phi^{x_\phi} F^{x_F}$ is chosen, with the size exponent $x_\phi$ and the flux exponent $x_F$ as free parameters.
The resulting $\dsigW$ is shown in the right panel of \figref{f:S2N_exponents}.
Again, the region for $x_\phi=0$ where $\dsigW$ is small is very noticeable.
Similarly to the optical survey, this is because the source density per flux interval closely follows a power law, which can be seen in the right panel of \figref{f:distributions}.
Varying $x_\phi$ and $x_F$, we find a maximum at around $x_\phi=-1$ and $x_F=0.4$. These are the exponents that we use in our analysis, for which $\dsigW=2.84$.
\begin{equation}
    W(\phi,F)=\phi^{x_\phi} F^{x_F} \, , \quad x_\phi=-1, \quad x_F=0.4 \, , \quad \text{where   } \,  \dsigW=2.84
    \, .
    \label{eq:exponents_SKA}
\end{equation}
The fact that the maximal $\Delta_W$ for SKA is  smaller than the one for Euclid/LSST already indicates that the latter may lead to a more accurate measurement of the velocity and the intrinsic dipole.

We also briefly explore the idea of weighting only by the flux \citep{Tiwari:2015_NVSS}.
This could be an advantage if a large fraction of the sources are not spatially resolved, i.e., the sizes have a large uncertainty and cannot be used in the analysis.
For the LSST/Euclid surveys, however, the case of weighting only by the magnitude is not as interesting since redshifts will be measured much more precisely than the sizes of small radio sources by the SKA.
For the SKA, setting $x_\phi=0$, we find a flat maximum of $\Delta_W \approx 0.4$ for $x_F\approx -1$.
For $x_F=1$, which corresponds to the dipole in sky brightness, $\Delta_W<0.1$ is even smaller.
Again, these values can possibly be optimized by also testing different flux cuts, which we do not explore here.
We compare the maximum possible $\dsigW=2.84$ when the sizes are used with the maximum of $\dsigW=0.4$ when no sizes are involved.
The signal-to-noise ratio \eqref{eq:signal_to_noise_beta} scales with $\sqrt{\Ntot}\Delta_W$. Therefore, to achieve the same precision without using sizes, one  needs $(2.84/0.4)^2 \approx 50$ times the number of sources when analysing with the sources' sizes.
In other words, as soon as more than a fraction $1/50$ of the sources are resolved well enough (for example $\sigma_\phi=0.1\, \arcsec$ and using sources with $\phi>0.3\, \arcsec$ as discussed in the next section), weighting by a combination of flux and size becomes a better choice than weighting by a power of the flux to measure $\vecbeta$.

\subsection{Results}
\label{s:results}
In this section, the results of our new method using the estimators \eqref{eq:estbeta} and \eqref{eq:estdint} with weight functions \eqref{eq:exponents_LSST} and \eqref{eq:exponents_SKA} applied to the LSST/Euclid and SKA simulations respectively, are discussed and compared to the method employing only the number count dipole.
We begin with the measurement of our peculiar velocity $\vecbeta$.

\begin{figure}
    \centering
    \includegraphics[width=\textwidth]{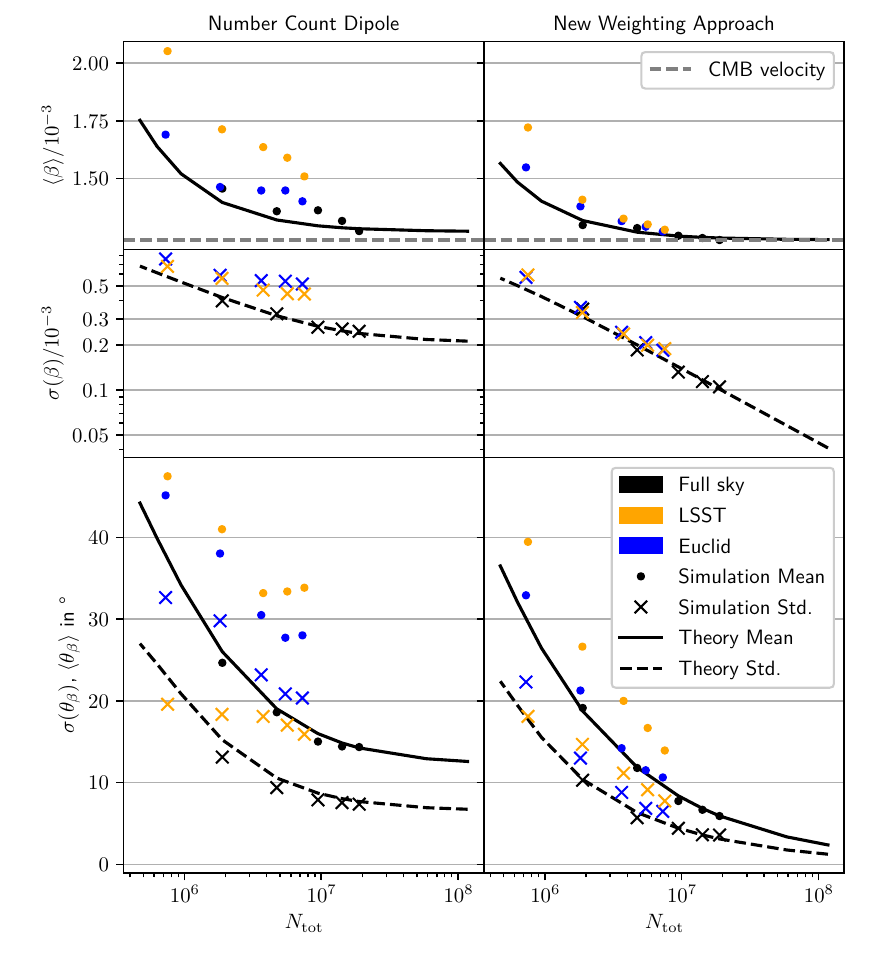}
    \caption{
    Estimating the velocity $\beta$ (top and middle row) and the deviation angle $\theta_\beta$ (bottom row), for different total number of sources $\Ntot$.
    Results from the number count method (left column) are compared to the new weighting method \eqref{eq:estbeta} (right column).
    $\beta$ and $\theta_\beta$ are estimated in $100$ mock catalogs for each $N_{\rm tot}$;  means and standard deviations of these results are shown here as dots and crosses respectively.
    Different colors indicate the sky coverage with which the analysis was performed.
    Solid and dashed lines indicate the theoretical mean and standard deviation respectively, calculated with \eqref{eq:mean_and_sigma_d}, \eqref{eq:mean_sigma_theta} (number count method) and \eqref{eq:mean_and_sigma_beta}, \eqref{eq:mean_and_sigma_theta_int_beta} (new weighting approach) and indicate the expectations for the behaviour of the full sky forecast. 
    Note that these results assume perfect measurements of magnitude and redshift.
    See main text for discussion.
    }
    \label{fig:LSST_results_beta_sources_used}
\end{figure}

\begin{figure}
    \centering
    \includegraphics[width=\textwidth]{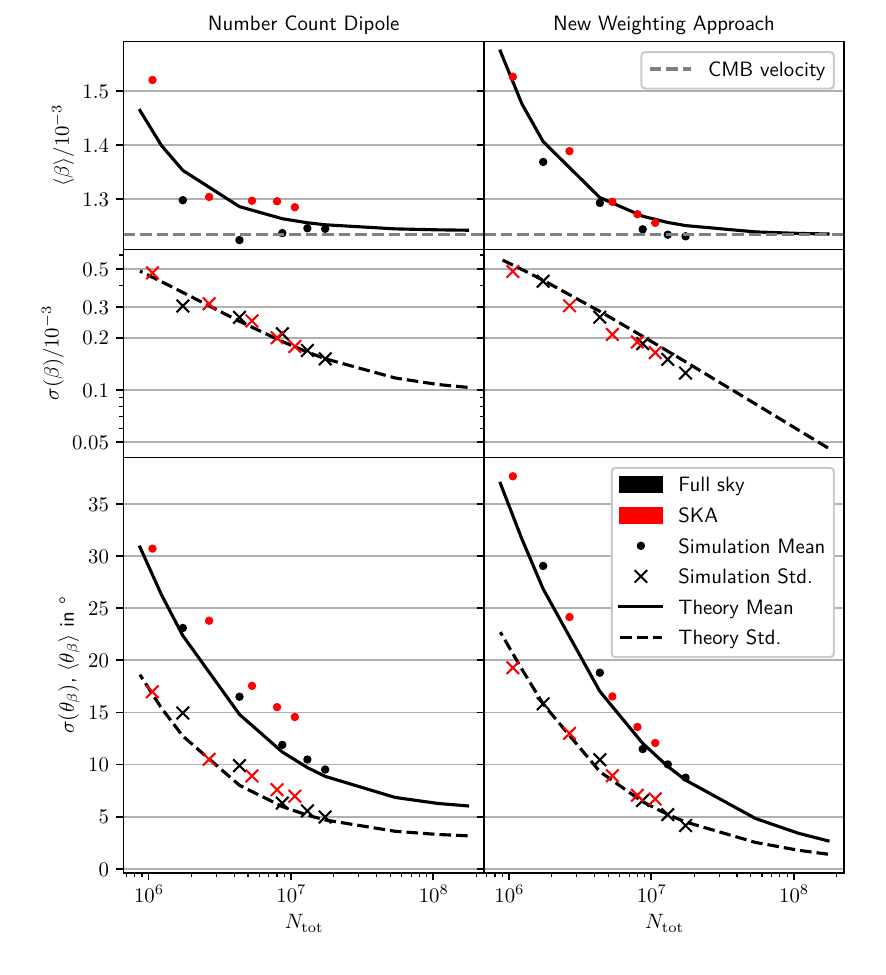}
    \caption{
    The same as \figref{fig:LSST_results_beta_sources_used} but for SKA mock catalogs as defined in \secref{s:applications}.
    Note that these results assume perfect measurements of flux and sizes.
    The inclusion of measurement errors is studied separately (see \figref{fig:new_LSST_SKA_errors}).
    }
    \label{fig:SKA_results_beta_sources_used}
\end{figure}

\paragraph{Estimating $\vecbeta$}
The results of estimating $\vecbeta$ are displayed for the LSST/Euclid in \figref{fig:LSST_results_beta_sources_used} and for the SKA in \figref{fig:SKA_results_beta_sources_used}.
These figures show the mean and standard deviation of the estimated velocities $\beta$ and the deviation angle $\theta_\beta$ found in $100$ mock catalog realizations.
These values reflect the bias and the uncertainty with which the speed and direction of our velocity can be measured.
Of course, the number count method does not estimate the velocity directly, but instead the combined dipole which includes the intrinsic dipole.
Nevertheless, to compare to the new weighting approach, the number count dipole amplitudes are divided by $\ampNC$ to give an estimate of the velocity.

To better understand the influence of shot noise, we show the results for a range of $\Ntot$, which for any sky coverage refers to the total number of sources in the observed region.
Due to computational effort of simulating each source individually, we are limited to $\Ntot \approx 2\times 10^{7}$.
Since we are able to extrapolate these results, this number of sources is sufficient.
We also compare the results to the expected bias and uncertainty for the full sky analysis.
We thus show the mean and standard deviation of both $\beta$ and $\theta_\beta$ calculated with \eqref{eq:mean_and_sigma_d} and \eqref{eq:mean_sigma_theta} in the case of the number count method.
For the new weighting approach, we show the results of \eqref{eq:mean_and_sigma_beta} and \eqref{eq:mean_and_sigma_theta_int_beta}. 
In Figs.~\ref{fig:LSST_results_beta_sources_used} and \ref{fig:SKA_results_beta_sources_used} one sees that the agreement between these calculations and the simulations on the full sky is excellent.

Let us first discuss the full sky results before focusing on the differences from partial sky coverage.
For both LSST and SKA results, the trend is naturally that the more sources are observed, the bias and uncertainty in the velocity and the deviation angle all decrease. Their measurements become more accurate and more precise.
However, there is a crucial difference between the number count method and the new weighting approach. 
In the former, the uncertainty of estimating the velocity, $\sigma_\beta$ (shown as dashed lines (theoretical curve) and black dots (simulations) in the top panels of Figs.~\ref{fig:LSST_results_beta_sources_used} and \ref{fig:SKA_results_beta_sources_used}) converges to a constant, whereas for the latter, the uncertainty continues to decrease like $1/\sqrt{\Ntot}$.
Similarly, the deviation angle (bottom panels of Figs.~\ref{fig:LSST_results_beta_sources_used} and \ref{fig:SKA_results_beta_sources_used}) tends much closer to zero when using the weighting method.

The reason for this improvement is that the combined number count dipole has two random contributions. While the first of them, the shot noise contribution, indeed goes to zero for large $\Ntot$, the second one, the intrinsic clustering dipole, remains.
Hence, even in the limit $\Ntot \ra \infty$, the number count dipole does not converge to the kinematic dipole, but has an unknown contribution from the random intrinsic dipole.
On the other hand, the weighting method separates the intrinsic dipole from the kinematic effect and is thus affected by shot noise only, which can be arbitrarily small when increasing $\Ntot$.
This allows us to measure our velocity amplitude and direction with a much higher precision, given enough sources.
This difference between the two methods is less pronounced in the SKA forecast than in the LSST or Euclid forecasts because the intrinsic dipole is larger in the latter (see right panel of \figref{fig:ang_power_spectrum_redshift}).

We now consider how results on the partial sky deviate from the full sky results in different ways.
Due to an incomplete sky coverage, there is a bias in the direction of the random dipoles as well as leakage from higher multipoles into the random dipole amplitudes.
This is true for the shot noise and for the intrinsic clustering.
The directional bias is the reason why the effect of incomplete sky coverage is larger when analysing the deviation angle than when analysing the velocity.
Since we keep the total number of sources distributed on the observed region of the sky fixed, the number density is larger for a smaller sky coverage, which partly cancels the shot noise leakage.
The total leakage effect is thus dominated by the intrinsic clustering.

This explains why for the weighting method, where the intrinsic dipole of the partial sky is removed, i.e. including the leakage from higher multipoles, the partial sky analyses (LSST, Euclid and SKA) all deviate less from the full sky analyses than for the number count method.
The weighting method is, as explained above, not at all affected by intrinsic clustering and therefore, the leakage effect does not play as an important role.
The directional bias from partial sky coverage is largest for the LSST mask, which covers only parts of one hemisphere.
Hence, the largest deviation from the full sky analysis
is seen in the LSST deviation angle.
Since the SKA covers the largest fraction of the sky,
the full sky analysis and the partial sky analysis of the SKA are very similar compared to the difference between full sky and the Euclid or even LSST sky coverages.

The excellent agreement between theory and full sky results validate extrapolating the results to larger $\Ntot$ which are computationally expensive to simulate.
For the cases of partial sky coverages, we can nevertheless estimate how they behave for larger $\Ntot$ by extrapolating the results shown Figs.~\ref{fig:LSST_results_beta_sources_used}~and~\ref{fig:SKA_results_beta_sources_used}.
To do so, we fit the expressions for mean and variance of the estimators presented in \secref{s:new_SN} to the results for the partial sky coverages with one free parameter that multiplies the shot noise variance.

We now compare our analysis of the number count dipole in the SKA with the forecast of Ref.~\cite{Bengaly_2019}.
Assuming $\Ntot = 3.3\times 10^{8}$ above $10\mu$Jy, in their forecast, a relative uncertainty of the number count dipole amplitude of $12\%$ is found. We find the same uncertainty of the velocity estimate of $\sigma_\beta/\beta \approx 12\%$.
In~\cite{Bengaly_2019}, the galactic longitude and latitude can be determined with an uncertainty of $13\degree$ and $6\degree$, respectively.
Combining these, this corresponds to a mean deviation angle of approximately $10\degree$.
Inspecting the trend of the mean deviation angles $\expect{\theta_\beta}$ towards larger $\Ntot$ (see \figref{fig:SKA_results_beta_sources_used}, bottom left),
this agrees well with our results for the SKA sky coverage.
Comparing these results to the weighting method, the same accuracy and precision can be achieved with approximately $4\times 10^7$ sources for which the size is perfectly measured.
If the sizes of all the $3.3\times 10^{8}$ were available, one could measure $\beta$ within approximately $\sigma_\beta/\beta \sim 2.5\%$ and its direction within $\expect{\theta_\beta} \sim 2.2\degree$.

Extrapolating the LSST (Euclid) forecast to $\Ntot=10^9$ sources, the velocity can be estimated with $\sigma_\beta/\beta \sim 1.4\%$ ($1.3\%$).
The mean deviation angle is $\expect{\theta_\beta} \sim 1.2 \degree$ ($0.9\degree$).
The uncertainty in $\beta$ is smaller than the systematic limit through the intrinsic dipole in the redshifts, for example the bulk flow, which is approximately $0.02\beta$ (see Appendix~\ref{app:bulk_flow}).
All these results are summarized in Table~\ref{t:summary}.

We note that this is a great improvement in comparison to the results with the number count dipole, where $\sigma_\beta/\beta$ converges to approximately $30\%$ and $\expect{\theta_\beta}$ to well above $20\degree$.
The estimate of the velocity through the number count dipole can be improved by removing the closest sources, i.e.~small redshift sources.
This is because clustering is largest on small scales (compare to \figref{f:intrinsic} and see e.g.~\cite{Tiwari:2016_Cluster_Simulation}).
Removing local sources is straightforward in LSST/Euclid because redshift measurements are available and is also proposed in \cite{Bengaly_2019} for the SKA survey that we consider.
If, however, there is a large intrinsic anisotropy which does not originate in clustering as expected in the $\Lambda$CDM model, only the new weighting approach allows to separate this from the kinematic dipole.

In the next section, we consider measurement uncertainties.
In particular, we will show that small enough measurement errors to not degrade these results.

\paragraph{Measurement errors}
Any measurement of the properties magnitude, flux, redshift and size has a measurement error.
To see how this affects the estimate of our velocity with the weighting method, we include these errors as described in \secref{s:applications}.
\ch{Exemplarily, for the two applications, a redshift uncertainty $\sigma_z$ and an uncertainty in the size estimate $\sigma_\phi$ are considered, respectively.
Uncertainties in flux and magnitude would affect both the number count dipole and the new weighting approach.
As we discuss at the end of this paragraph, measurement errors of these two quantities are small and are hence expected to have no relevant effect on our method.
It is therefore sufficient to only study redshift and size errors.
}
In the LSST/Euclid (SKA) simulations, after including the kinematic effects, each redshift $z$ (size $\phi$) is replaced by a Gaussian random variable with mean given by the original value and standard deviation $\sigma_z$ ($\sigma_\phi$).
To avoid negative redshifts, we take the absolute value of the random redshift.
It is important for the analysis that the telescope does not observe differently in any direction.
Hence, we assume that neither the uncertainty nor the bias (which we take to be zero in general) is direction dependent.
In principle, a direction dependent sensitivity could be included in the mask, but this goes beyond the present `proof of principle' study.

\begin{figure}
    \begin{minipage}{0.5\textwidth}
        \centering
        \includegraphics[width=\textwidth]{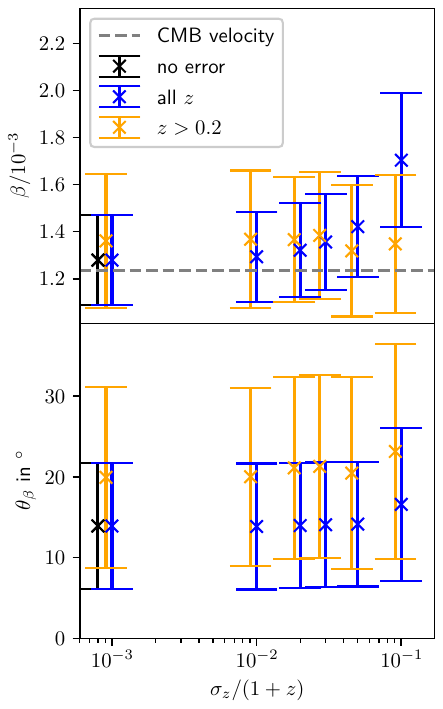}
    \end{minipage}\hfill
    \begin{minipage}{0.5\textwidth}
        \centering
        \includegraphics[width=\textwidth]{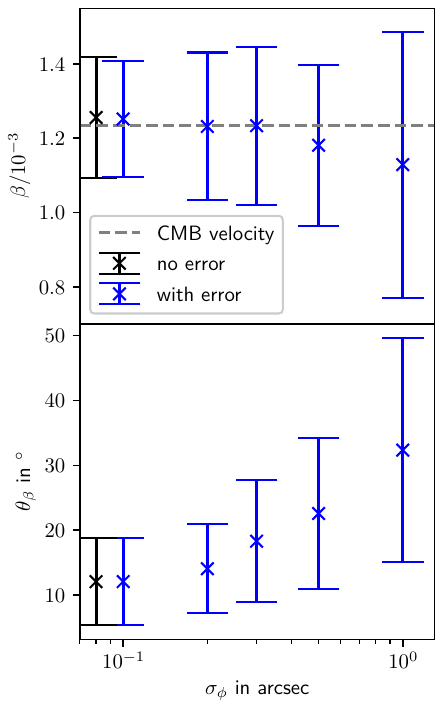}
    \end{minipage}
    \caption{\label{fig:new_LSST_SKA_errors} 
    Effect of measurement uncertainties on accuracy and precision of the velocity estimator \eqref{eq:estbeta}.\\
    \textit{Left:}
   LSST simulations: a redshift error $\sigma_z=(10^{-3} \sim 10^{-1}) (1+z)$ has been imposed on the mock catalogs, as described in the main text. For comparison, the case with zero measurement uncertainty (corresponding to the results for the largest $\Ntot=0.75 \times 10^7$ for LSST displayed in 
   \figref{fig:LSST_results_beta_sources_used}) is shown in black. Error bars show the mean and standard deviation of the velocity $\beta$ and deviation angle $\theta_\beta$ estimated in $100$ mock distributions.
    The blue error bars show the results when including all sources independent of their redshift (as done in \figref{fig:LSST_results_beta_sources_used}).
    The orange error bars show the results when only using sources with redshift larger than $0.2$ for the analysis. For better visibility they are slightly displaced toward the left.
    \newline
    \textit{Right:}
    Same as left panel, but for the SKA simulations with $\Ntot=1.1\times 10^7$: a measurement error in the size $\sigma_\phi$ has been included.
    The black error bar corresponds to the result without measurement uncertainties, which is also displayed in \figref{fig:SKA_results_beta_sources_used}.
    Sources with sizes $\phi>0.3\, \arcsec$ are used (see Equation~\eqref{eq:cut_SKA}).
    }
\end{figure}

The results including measurement errors are shown in \figref{fig:new_LSST_SKA_errors}.
This time, we show the mean and standard deviation of the velocity and the deviation angle as one error bar for fixed $\Ntot=0.75 \times 10^7$ (LSST) and $\Ntot=1.1\times 10^7$ (SKA).
For comparison, the outcome without any measurement uncertainties is shown in black.
We use the LSST and SKA sky coverages respectively, but this does not influence the result.
For significantly large measurement uncertainties, the error in the results increases, and the results also become biased.
This is because calculating the kinematic dipole amplitudes does not give correct results if the observed source density $n(F,\phi)$ or $n(m,z)$ differs significantly from the rest frame density due to measurement errors, i.e., the amplitudes of the kinematic dipoles $\ampNC$ and $\ampW$ become biased.

For the LSST/Euclid forecast, we also test imposing a cut of $z>0.2$.
Since the error is proportional to $(1+z)$, the relative error is largest for small redshifts.
Therefore, this cut suppresses the bias. However, the scatter increases because less sources are available for the analysis.
We note that with this cut, we do not attempt to suppress the clustering itself but only the sources whose redshifts have the largest relative uncertainty.
We now see that measurement uncertainties below $\sigma_z/(1+z) = 5\% $ do not degrade the results (see orange error bars in \figref{fig:new_LSST_SKA_errors}).
This explains the choice of the photometric survey of Euclid rather than the spectroscopic one.
The latter will observe much less sources, so the statistical uncertainty is larger, while the higher redshift resolution does not improve the result.

For the SKA forecast, where all sources with a size $\phi>0.3\, \arcsec$ are used, we conclude that an error of $\sigma_\phi \sim 0.1\,  \arcsec$ does not affect the results.
Applying a more stringent cut might also decrease the bias but it would also greatly increase the shot noise, since there are many sources with $\phi\sim  0.3 \, \arcsec$ (see \figref{f:distributions}).
For our model of the source distribution, the cut $\phi>0.3 \, \arcsec$ reduces the total number of observable sources to a third.
Therefore, in addition to the forecast with the total number of expected sources $\Ntot=3.3\times 10^8$~\cite{Bengaly_2019}, we also consider the realistic scenario of $\Ntot=10^8$ well resolved sources in the summary Table~\ref{t:summary}.

\begin{figure}
    \centering
    \includegraphics[width=\textwidth]{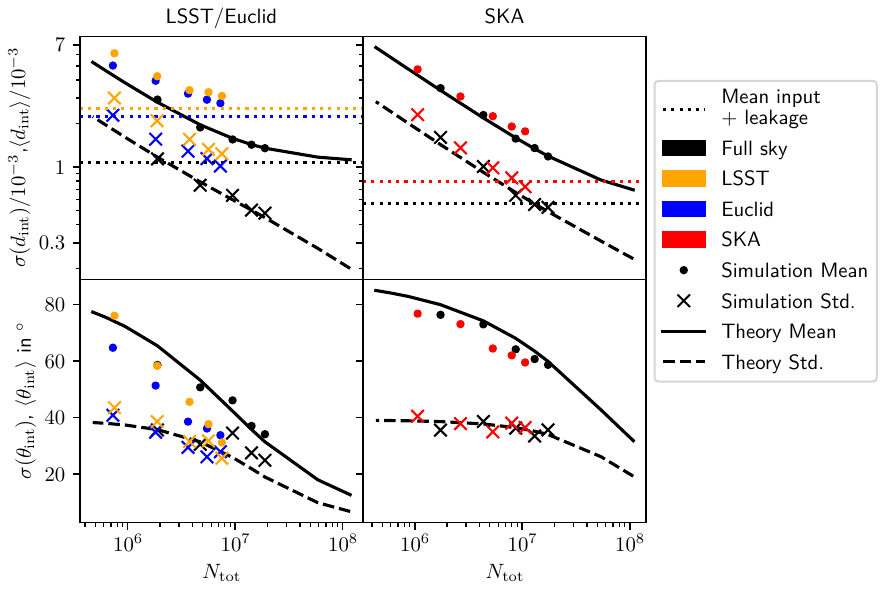}
    \caption{
    Accuracy and precision of the estimator for amplitude and direction of the intrinsic dipole with the new weighting approach \eqref{eq:estdint} for LSST/Euclid (\textit{left}) and SKA (\textit{right}).
    The top panels show the mean of estimating the intrinsic dipole amplitude and its standard deviation when compared to the true clustering for each mock distribution.
    The bottom panels show the mean and standard deviation of estimating the deviation angle, which indicates how well the direction of the intrinsic dipole amplitude can be inferred.
    The dotted lines show the mean intrinsic dipole amplitude present in the simulations.
    They include the expected leakage due to partial sky coverage, i.e., higher clustering multipoles lead to an increased estimated intrinsic dipole amplitude which for the cases LSST and Euclid actually doubles the intrinsic dipole.
    Solid and dashed lines indicate the theoretical mean and standard deviation of the estimator, respectively, calculated with \eqref{eq:mean_and_sigma_int} and \eqref{eq:mean_and_sigma_theta_int_beta}. 
    }
    \label{fig:new_results_intdip}
\end{figure}

The main results of this work, Figures~\ref{fig:LSST_results_beta_sources_used}, \ref{fig:SKA_results_beta_sources_used} and \ref{fig:new_results_intdip}, which are summarized in Table~\ref{t:summary}, are thus valid for $\Ntot$ being the total number of well resolved sources, which corresponds to $\sigma_z/(1+z) \lesssim 5\%$ while using sources with $z>0.2$ (LSST/Euclid), and $\sigma_\phi \sim 0.1\, \arcsec$ while using sources with $\phi>0.3\, \arcsec$ (SKA).

\ch{
We briefly come back to measurement errors in magnitude and flux. For LSST and Euclid, we expect all sources to have a high signal to noise ratio larger than 20 (gold standard)~\cite{LSST_2009}. For the SKA, the uncertainty in the flux is expected to be $\sigma_F \sim 0.1 \mu \mathrm{Jy/beam}\lesssim 0.01 F_\mathrm{min}$~\cite{Bengaly_2019}. Comparing this to the right panel of  \figref{fig:new_LSST_SKA_errors}, since magnitudes/fluxes enter similarly into the analysis as sizes (see Eqs.~\eqref{eq:exponents_LSST} and \eqref{eq:exponents_SKA}), such uncertainties do not bias the results or increase their errors by a relevant amount.
Hence, for these scenarios, only considering the larger errors in redshift and size is sufficient.
}

\paragraph{Estimating $\dipole_\mathrm{int}$}
We also test how well the intrinsic dipole can be measured using the estimator~\eqref{eq:estdint}.
To do so, we compare the estimated intrinsic dipole of each simulated mock with the respective true dipole.
The true intrinsic clustering is different for each mock distribution, as each is a random realization of the angular power spectrum.
To test accuracy and precision with which the intrinsic dipole can be measured with our estimator, we compare the amplitude and direction of the estimated value with the respective true values of each simulated mock distribution.

Furthermore, to take into account the large amount of leakage from anisotropies on smaller scales, we compare the estimated intrinsic dipole with the true dipole generated by clustering on all scales.
This is obtained by recording the dipole during the process of creating the mock catalogs after realizing a Gaussian field from the angular power spectrum and applying the mask.
Thereby, we obtain the true intrinsic dipole plus leakage, $\dipole_\mathrm{int}^\mathrm{t}$, for the particular mask geometry, which we compare to the estimated dipole that also includes the leakage $\dipole_\mathrm{int}$.
The mean of the estimated intrinsic dipole amplitudes $\expect{d_\mathrm{int}}$, the standard deviation $\sigma(d_\mathrm{int})$ of the differences of amplitudes $d_\mathrm{int}^\mathrm{t} - d_\mathrm{int}$, and the mean $\expect{\theta_\mathrm{int}}$ and standard deviation $\sigma(\theta_\mathrm{int})$ of the angles between the true and estimated directions $\theta_\mathrm{int}=\arccos (\myvec{d}_\mathrm{int}^\mathrm{t}\cdot \myvec{d}_\mathrm{int}/d_\mathrm{int} d_\mathrm{int}^\mathrm{t})$ are shown in \figref{fig:new_results_intdip}.
They are compared to the full sky expectations \eqref{eq:mean_and_sigma_int} as well as \eqref{eq:mean_and_sigma_theta_int_beta}.

This comparison is relevant to test how well our new approach works to compare the $\Lambda$CDM model with observations.
The $\Lambda$CDM expectation of the intrinsic clustering dipole for a given mask geometry can be obtained by realizing fields from the angular power spectrum, masking them, and then estimating the dipole.
Repeating this procedure many times, one obtains the expected clustering dipole on the particular mask,
which should be compared to the observations when our estimator of $\dipole_\mathrm{int}$ \eqref{eq:estdint} is applied to data to test the isotropy of the Universe.
Due to cosmic variance and the mask, the theoretical expectation contains a lot of variance because of which constraints on the large scale anisotropy are weaker.

To discuss \figref{fig:new_results_intdip}, we start with the full sky results.
First, we notice that they agree well with the theoretical results given in \eqref{eq:mean_and_sigma_int} and \eqref{eq:mean_and_sigma_theta_int_beta} which are indicated as solid and dashed lines in the figure.
In particular, the standard deviation $\sigma(d_\mathrm{int})$ decreases like $1/\sqrt{\Ntot}$ and the mean deviation angle $\expect{d_\mathrm{int}}$ also approaches zero for large $\Ntot$.
Comparing to the velocity estimates discussed above, the intrinsic dipole estimates are more biased and less precise.
This is because the  shot noise in  $\CintSN$ is larger than the shot noise in the velocity $\CbetaSN$
(compare \eqref{eq:C_beta} with \eqref{eq:C_int} with $\Delta_W=2.84$ or $\Delta_W=3.91$ and $\ampNC \approx 3.75$).
Even so, given a large number of sources and a large enough sky coverage, it possible to detect the intrinsic dipole to a good precision.

We turn our focus to the effect of incomplete sky coverage.
In an isotropic Universe with random clustering we expect similar amplitudes of all the low intrinsic multipoles.
Empirically, we therefore expect that the leakage enhances the intrinsic dipole by approximately a factor of $\gtrsim \sqrt{1/\fsky}$, which in this case of LSST/Euclid is $\gtrsim  \sqrt{2.5}$.
This shows why the mean intrinsic dipoles plus leakage (colored dotted lines in \figref{fig:new_results_intdip}) are larger by approximately a factor of $2$.
For the SKA forecast, the mean intrinsic dipole deviates not as much from the full sky results,
because the sky coverage of the SKA survey is much larger.
Note that for our velocity estimator, the effective intrinsic dipole is removed so that neither the intrinsic dipole itself nor any contributions from anisotropies in the number density on smaller scales are present.

While the intrinsic dipole by itself is not detectable because of the incomplete sky coverage, the intrinsic dipole plus leakage, can be estimated to arbitrary precision given enough sources, as we show in \figref{fig:new_results_intdip}.
The uncertainty $\sigma(d_\mathrm{int})$ as well as the mean deviation angle (and its scatter) all approach zero for large sources.
Extrapolating the partial sky results to $\Ntot=10^9$, we find for the LSST (Euclid) that the relative uncertainty with which $d_\mathrm{int}^\mathrm{t}$ can be measured is $\sigma(d_\mathrm{int})/d_\mathrm{int}^\mathrm{t} \sim 4.6\%$ ($4\%$) and the mean deviation angle is $\expect{\theta_\mathrm{int}}\sim 3.1\degree$ ($2.7\degree$).
For the SKA forecast, assuming $3.3\times 10^8$ ($10^8$) well resolved sources, we find $\sigma(d_\mathrm{int)}/d_\mathrm{int}^\mathrm{t} \sim 23\%$ ($39\%$) and $\expect{\theta_\mathrm{int}}\sim 13\degree$ ($24\degree$).
Here, the SKA forecast is significantly less precise also because the intrinsic clustering dipole is expected to be smaller.
These results, too, are summarized in Table~\ref{t:summary}.
Though we do not indicate the influence of measurement errors on the intrinsic dipole, the effects are similar to those discussed for the velocity estimator.
Again, we stress that all these results hold for $\Ntot$ referring to the total number of well resolved sources as defined in the previous paragraph.

To summarize, we have shown how the intrinsic dipole can be measured independently of our velocity.
Due to incomplete sky coverage and hence leakage from higher clustering multipoles, the clustering dipole is measured as an effective clustering for the particular mask geometry.

\section{Conclusions}\label{s:con}

In this paper, we have presented a novel method that disentangles the intrinsic dipole in the observed distribution of galaxies from the kinematic effects arising from our peculiar motion.
The intrinsic dipole represents the largest-scale anisotropy of the galaxy distribution in the mean rest frame of the galaxies.
In the full sky it is the dipolar anisotropy, but in a partial sky it also contains contributions from higher multipoles.
The kinematic effects are due to our motion with respect to that frame.

Our new approach relies on including other measured properties of galaxies that are also affected by the relative motion, such as their size and redshift.
By weighting each source by a combination of such properties, we can measure weighted dipoles in addition to the dipole in the number density.
Since kinematic effects enter differently, we combine the number count dipole with an optimally constructed weighted dipole to independently measure our velocity and the intrinsic dipole of the source distribution.
To do so, we argue that large scale anisotropies in the distribution of the quantities by which we weight the sources are much smaller, $\sim 10^{-5}$, than the velocity and intrinsic dipole which we aim to measure, which are $\sim 10^{-3}$.

We have shown how this approach can be employed in future galaxy surveys such as LSST or Euclid and in the upcoming radio survey by the SKA.
Using optimized estimators applied to mock catalogs of future surveys, we have forecasted the precision with which the velocity and the amplitude of the intrinsic dipole (including leakage from anisotropies on smaller scales) along with their directions can be measured. Our results are summarized in Table~\ref{t:summary}.

\begin{table}
\centering
\begin{tabular}{lccccc}
 &                                                            & LSST   & Euclid & SKA (realistic) & SKA (high resolution)     \\ \toprule
\multirow{4}{*}{Obs.~parameters} & $N_\mathrm{tot}$  & $10^9$ & $10^9$ & $10^8$ & $3.3 \times 10^8$  \\
 & $\sigma_z/(1+z)$ and $\sigma_\phi$                             & 5\%    & 5\%  & $0.1 \, \arcsec$  & $0$          \\
 & $z_\mathrm{min}$ and $\phi_\mathrm{min}$                             & $0.2$    & $0.2$  & $0.3 \, \arcsec$  & $0$           \\
 & $\fsky$                                                    & $40\%$ & $38\%$ & $61\%$  & $61\%$  \\ \midrule
\multirow{4}{*}{Forecast}                 & $\sigma(\beta)/\beta$   &  $1.4\%$      & $1.3\%$ & $4.5\%$ &   $2.5\%$             \\
                     & $\langle \theta_\beta \rangle$        &    $1.2 \degree$ & $0.9\degree$ & $3.9\degree$ &  $2.2\degree$  \\
                     & $\sigma(d_\mathrm{int})/d_\mathrm{int}^\mathrm{t}$   &      $4.6\%$ & $4\%$ & $39\%$ & $23\%$         \\
                     & $\langle \theta_\mathrm{int} \rangle$&    $3.1 \degree$ & $2.7\degree$ & $24\degree$ &  $13\degree$ \\ \bottomrule
\end{tabular}
    \caption{Summary of results obtained in this work.
    The first set of four rows shows the expected observational parameters: the total number of observed sources $N_{\rm tot}$, the maximum measurement error of the redshift $\sigma_z/(1+z)$ or the size $\sigma_{\phi}$, the lower redshift and size cuts $z_\mathrm{min}$ and $\phi_\mathrm{min}$ and the sky coverage $f_{\rm sky}$.
    The second set of four rows shows the significance of detection of the observables: amplitude and direction of our velocity and the intrinsic dipole.
    These forecasts are displayed in four columns for LSST, Euclid, and the SKA, respectively.
    The values are obtained by extrapolating the simulation results (Figures \ref{fig:LSST_results_beta_sources_used}, \ref{fig:SKA_results_beta_sources_used} and \ref{fig:new_results_intdip}) to larger $\Ntot$.
    We assume that there are $\Ntot$ well resolved sources, i.e.~sources for which $z>z_\mathrm{min}=0.2$ and $\sigma_z/(1+z)<5\%$ (which is the expected photometric resolution for LSST~\cite{LSST_2009} and Euclid~\cite{Euclid_def_study}) and $\sigma_\phi=0.1\, \arcsec$ and $\phi>\phi_\mathrm{min}=0.3\, \arcsec$ for all sources in SKA (realistic).
    Additionally, we show the results for an ideal high resolution SKA survey, where the sizes of all $3.3 \times 10^8$ sources are perfectly measured.
    The SKA observes the intrinsic dipole less precisely not only because there are less sources but also because we show the relative precision here, and the expected clustering dipole is smaller.
    We note that the systematic uncertainty of $\approx 2\%$ in $\beta$ and $d_\mathrm{int}^t$ is not reflected in the errors reported here.
    }
    \label{t:summary}
\end{table}
For LSST and Euclid-like surveys, we assumed $\Ntot=10^9$ sources with redshifts $z>z_\mathrm{min}=0.2$ and a maximum redshift uncertainty of $\sigma_z/(1+z)=0.05$.
For the SKA, we assumed two scenarios, a realistic one with $\Ntot=10^8$ well resolved sources, i.e., $\phi>\phi_\mathrm{min}=0.3\, \arcsec$ and $\sigma_\phi=0.1\, \arcsec$, as well as a second ideal high resolution scenario where the sizes of all $\Ntot=3.3\times 10^8$ sources above $10\mu$Jy are well measured.
We find that the present approach allows to measure our velocity to a precision $\sigma _\beta/\beta \sim 1.3 - 4.5\%$ and the direction of our motion to $\theta_{\beta} \sim 0.9\degree - 3.9\degree$.
The LSST and Euclid surveys should be able to measure the intrinsic dipole anisotropy plus leakage to $\sigma(d_{\rm int})/d_{\rm int}^\mathrm{t}\lesssim 5\%$.
In fact, one is able to reach the limit of the systematic uncertainty of approximately $2\%$ imposed by the effect of $\Lambda$CDM perturbations on the weights.

These results have far-reaching implications. Measuring our velocity with respect to the mean rest frame of distant objects independently of their number count anisotropy yields important information.
Within $\Lambda$CDM, this value should agree with the velocity inferred by the CMB dipole and comparing the two is an important test of the Cosmological Principle, a fundamental building block of the standard model of cosmology.
The approach presented in this paper allows to determine an independent value of our velocity with respect to the frame defined by the observed sources. Knowing this velocity is crucial to determine the frame in which large-scale structure data or supernovae data should be analysed, e.g., to determine the correct reference frame for the possibility of anisotropic cosmic expansion as advocated in~\cite{sarkar_cosmic_acceleration}.

Measuring the intrinsic large scale matter anisotropy is equally important.
In \cite{Secrest_2021}, an excess of the dipole amplitude in the distribution of quasars in comparison to the prediction of the $\Lambda$CDM model was announced with $4.9\sigma$ significance.
Apart from unaccounted for systematics, there are two possible explanations for this:
(i) either the average reference frame of cosmic galaxies is not at rest with respect to the surface of last scattering, i.e.,~the reference frame where the CMB dipole vanishes as mentioned above, or (ii) there is a large unexpected intrinsic anisotropy in the matter distribution traced by the quasars. Both these interpretations are in tension with the Cosmological Principle.
Analysing only the number count dipole does not allow us to distinguish which of the above two $\Lambda$CDM expectations is violated.
The new approach described in this paper allows to do so and therefore serves as an important test of fundamental assumptions in the $\Lambda$CDM model.

There are several directions in which this work can be taken further.
For example, one might consider multiple flux measurements of each source at different frequencies as an additional property.
Since the spectral index is not constant over the full frequency range, and the Doppler boost depends on the spectral index, the kinematic contributions also enter differently in weighted dipoles constructed with flux measurements at different frequencies.
From our analysis in the present paper, this, or any other measured source property which transforms in a well defined way under boosts, is able to yield measurements of both, the intrinsic dipole of the source distribution and our velocity with respect to the mean rest frame of the sources independently. It may also be interesting to investigate the potential of weighted multipoles instead of only a weighted dipole.
In this paper we concentrated on extracting the observer velocity, which dominates in the dipole. Applying the weighting scheme to higher multipoles may allow us to measure other quantities that dominate on smaller angular scales.

\acknowledgments
We thank Carlos Bengaly, Charles Dalang and Subir Sarkar for useful discussions.
We also thank the anonymous referee for a helpful report.
TN acknowledges support from Jochen Weller and financial support from the Swiss-European Mobility Programme and the Cusanuswerk. RD and MK  acknowledge funding from the Swiss National Science Foundation (SNSF). HP acknowledges support from the SNSF via Ambizione grant PZ00P2\_179934.
We acknowledge the use of NumPy~\cite{numpy}, SciPy~\cite{scipy}, Astropy~\cite{astropy1,astropy2}, healpy~\cite{healpy1}, HEALPix~\cite{healpy2} and Mathematica~\cite{mathematica}.
Plots have been produced with Matplotlib~\cite{matplotlib}.

\appendix

\section{General relativistic number count dipole}
\label{app:gr_kinematic}

Here, we explicitly show that a fully general relativistic expression of the kinematic dipole is equivalent to the special relativistic expression to first order in the absence of redshift information.
We start with the formula for the general relativistic redshift dependent amplitude of the kinematic dipole in the observed galaxy distribution given in \cite{Maartens:2017qoa} (see also 
\cite{Bonvin:2011bg,Challinor:2011bk,DiDio:2013bqa,Scaccabarozzi_2018,Bertacca:2019wyg})
\begin{equation}
\dip(z)=
\Big[
3
+\frac{\dot{H}}{H^2}
+(2-5s)\frac{1+z}{rH}
-\fevo
\Big]\beta
+\order{2}
\label{eq:full_dipole_formula}
\end{equation}
with the evolution bias
\begin{equation}
\fevo({z},F)=- \frac
{\partial \ln [(1+{z})^{-3}\Ncurl({z},>F)] }
{\partial \ln (1+{z})}
\label{eq:fevo_definition}
\end{equation}
and the magnification bias
\begin{equation}
s(\bar{z},F)=\frac
{\del \log_{10} \Ncurl (\bar{z},<m) }
{\del m}
= - \frac{2}{5} \partialder{\ln \Ncurl (\bar{z},>F)}{\ln F}
\label{eq:magnbias_def}
\, .
\end{equation}
Only first order terms are included.
A dot refers to a derivative with respect to physical time and $r$ is the comoving radial distance. $a=(1+z)^{-1}$ is the scale factor and $H=\dot{a}/a$ the Hubble parameter. ${\Ncurl(z,>F)=\int_{L(z,F)}^\infty \mathrm{d}L' n_s(z,L')}$ is the background comoving number density of sources with flux exceeding $F$. It is given by integrating the luminosity function $n_s(z,L')$ over the range of luminosities that are observed with a flux larger than $F$.
A perfect flux-limited telescope would observe ${\mathrm{d}N/(\mathrm{d}\Omega\mathrm{d}z) \defined}\, {\nbar(z,>F) =}\, {(r^2 a^3 /H) \Ncurl(z,>F)}$ sources per solid angle and per redshift.
We rewrite the evolution bias in terms of $\nbar$, also making the dependence on the redshift explicit by introducing partial derivatives with respect to $z$ and $L$
\begin{align}
    \fevo(z,F)
    &=
    \frac{\dot{H}}{H^2}
    +\frac{2}{arH}
    -\partialder{\ln \bar{n}(z,>F)}{\ln (1+z)}
    \\
    &=
    \frac{\dot{H}}{H^2}
    +\frac{2(1+z)}{rH}
    -\der{\ln \bar{n}(z,>F)}{\ln (1+z)} 
    +\partialder{\ln \bar{n}(z,>F)}{\ln L} \partialder{\ln L}{\ln (1+z)}
    \, .
    \label{eq:fevo_rewritten}
\end{align}
One recognizes the magnification bias $s(z) = -\frac{2}{5}\partialder{\ln \Ncurl}{\ln L}$ since at fixed resshift, a derivative w.r.t.~the logarithm of the flux is equal to one w.r.t.~the logarithm of the luminosity \cite{DiDio:2013bqa}.
A flux cut corresponds to a cut in luminosity, which depends explicitly on the redshift via
\cite[equation (B11)]{Alonso:2015uua}
\begin{equation}
L_\mathrm{}(z)=4\pi F
(1+z)^{1+\alpha} r^2
\, ,
\end{equation}
where a power-law in the frequency dependence of the luminosity $L \propto \nu^{-\alpha}$ is assumed.
We find the explicit redshift dependence of the luminosity
\begin{equation}
\partialder{\ln L}{\ln (1+z)}=\frac{2(1+z)}{rH}+(1+\alpha)
\, .
\end{equation}
Using this in \eqref{eq:fevo_rewritten}, we find for the dipole amplitude in \eqref{eq:full_dipole_formula}
\begin{equation}
    \dip(z)=
    \Big[
    3
    +x(1+\alpha)
    +\der{\ln \bar{n}(z,>F)}{\ln (1+z)}
    \Big] \beta
    \, .
\end{equation}
Here, we reintroduced the slope $x$ of the cumulative number of sources at the flux limit, $N (> F) \propto F^{-x}$, which is related to the magnification bias by $s=x/2.5$.
Averaging over the full redshift range, the dipole amplitude of the projected source density is
\begin{equation}
    \dip =
    \frac{1}{N}\int_0^\infty  \mathrm{d}z  \,
    \dip(z) \nbar(z,>F)
    =[2+x(1+\alpha)]\beta
    \, .
\end{equation}
This assumes that $\expect{\alpha(z)x(z)}_z = \expect{\alpha(z)}_z \expect{x(z)}_z$. It is used that $n(z)$ as well as $z \times n(z)$ vanish for both $z\ra 0$ and $z \ra \infty$. $N$ is the total number of sources, $N=\int_0^\infty \mathrm{d}z\, \nbar(z,>F)$.

With this, we have found the equivalence of the special relativistic result for the kinematic dipole amplitude \eqref{eq:EllisBaldwinResult} of~\cite{1984MNRAS.206..377E} and the corresponding general relativistic expression \eqref{eq:full_dipole_formula} of~\cite{Maartens:2017qoa}.
To be more precise, we have shown that there are no new leading order contributions from the fully general relativistic treatment of the kinematic dipole amplitude when averaged over the full redshift range.
This short calculation can serve as a starting point for the inclusion of second order corrections to the kinematic dipole.

\section{\ch{The intrinsic dipole}}\label{app:Dip-int}
\ch{ In this appendix we briefly introduce the dipole from clustering in the standard $\La$CDM model. More details on the observable large scale clustering and its angular power spectrum can be found in~\cite{Bonvin:2011bg,Challinor:2011bk}. The two point correlation function is studied in \cite{Scaccabarozzi_2018,Bertacca:2019wyg,Tansella:2018sld}.  Within linear perturbation theory, which is very good on the large scales relevant for the dipole, and taking into account that our measurements are made on our perturbed background lightcone, one finds the following expression for the galaxy number count fluctuations at a fixed redshift $z$ (see~\cite{2009PhRvD..80h3514Y,Bonvin:2011bg,Challinor:2011bk,Camera:2014bwa,Jelic-Cizmek:2020pkh}):
\bea
\Delta(z, \mathbf{n})&=&b(z) \delta+\frac{1}{\mathcal{H}}\partial_r^2V
+\left(5s(z)-2\right)\int_0^{r} \mathrm dr' \frac{r-r'}{2rr'}\Delta_\Omega(\Phi+\Psi) \nonumber\\
&&-\left(1-5s(z)-\frac{\dot{\mathcal{H}}}{\mathcal{H}^2}+\frac{5s(z)-2}{r\mathcal{H}} +f_{\rm evo}(z) \right)\partial_rV-\frac{1}{\mathcal{H}}\partial_r\dot{V}+\frac{1}{\mathcal{H}}\partial_r\Psi\nonumber\\
&&+\frac{2-5s(z)}{r}\int_0^{r} \mathrm dr'(\Phi+\Psi)+(f_{\rm evo}(z)-3)\mathcal{H}V+\Psi+(5s(z)-2)\Phi\nonumber\\
&&+\frac{1}{\mathcal{H}}\dot{\Phi}+\left(\frac{\dot{\mathcal{H}}}{\mathcal{H}^2}+\frac{2-5s(z)}{r\mathcal{H}}+5s(z) -f_{\rm evo}(z) \right)\left[\Psi+\int_0^{r}  \mathrm dr'(\dot{\Phi}+\dot{\Psi})\right]\, . \label{e:DezNF}
\eea
Here $\de$ is the matter density contrast in comoving gauge, $V$ is the velocity potential in longitudinal gauge and $\Phi$ and $\Psi$ are the so called Bardeen potentials. All these functions are to be evaluated at the spatial position $\bx=r(z)\bn$ and time $t(z)=t_0-r(z)$, where $r=r(z)$ is the comoving distance out to redshift $z$ and $t_0$ is present time. $\HH$ is the conformal Hubble parameter and the operator $\De_\Omega$ denotes the Laplacean on the 2-sphere.
The functions $b(z)$, $s(z)$ and $f_{\rm evo}(z)$ are the galaxy bias, the magnification bias and the evolution bias, respectively. They depend on the specifics of the galaxy survey (which types of galaxies are observed)  and on the instrument (flux limit, frequency band).} 

\ch{The first term on the right hand side of~\eqref{e:DezNF} is the naively expected density fluctuations while the second term is the well-known redshift space distortion~\cite{1987MNRAS.227....1K,Hamilton:1995px} (see also~\cite{Bertacca:2019wyg}). The last term on the first line comes from lensing convergence which on the one hand reduces the density by enhancing the angular size of a given patch in the sky (pre-factor $-2$) but on the other hand enhances the number of galaxies seen above a fixed flux limit (pre-factor $+5s(z)$). We have found that the lensing contribution to the dipole is very small. This is not so surprising as lensing is the integrated effect from many smaller scale fluctuations and is therefore subdominant on large scales. The other terms come from the redshift due to the gravitational potential, the Shapiro time delay, the integrated Sachs Wolfe term and more, as detailed in~\cite{Bonvin:2011bg,Challinor:2011bk,Camera:2014bwa}.}

\ch{In order to find the result for a given redshift distribution $n(z)$ of galaxies, one has to integrate the above expression multiplied by $n(z)$ over redshift. There are public codes to determine the angular power spectrum of these fluctuations and we used the most recent version of CLASSgal~\cite{DiDio:2013bqa} which is included in the distribution {\sc class}~\cite{Blas:2011rf}.  The expression given in \eqref{e:DezNF} neglects terms at the position of the observer. However, these terms include, apart from the kinematic Doppler term which we treat separately, only monopole contributions and are therefore irrelevant for our analysis of the dipole.}

\ch{
On small scales, one expects only the first three terms to be relevant since the others are suppressed by factors $\HH/k$ and $(\HH/k)^2$, where $k$ is the wave number of the perturbation. However, we have found that the most relevant contributions  of the above expression also to the dipole come from the first two terms.
If Fig.~\ref{f:intrinsic} we show the full result (solid) and the result without redshift space distortion (dashed) which is dominated by the density term. Contrary to the bias, our choices of $s(z)=0$ and $f_{\rm evo}(z)=0$ are therefore not very relevant. 
}

\section{\ch{Mean and variance of the non-central $\chi(3)$ distribution}}
\label{app:derivation_non_central}
\ch{
In this appendix, we derive equation \eqref{eq:mean_and_sigma_d} making use of Refs.~\cite{noncentral_chi2,Nuttall_QM}.
The goal is to obtain an expression for the mean of the dipole amplitude $d$, which follows a non-central $\chi(3)$-distribution.
First, we introduce the components $\tilde{d}^{x/y/z}$ normalized by their standard deviation $\sigma_r$ to be Gaussian variables with variance $1$,
\begin{equation}
    \expect{\tilde{d}^{x/y}}=0 \, , \quad \expect{\tilde{d}^{z}}=d_\mathrm{kin}/\sigma_r
    \, , \quad 
    \tilde{d}^2= \sum_{i=\{x,y,z\}} (\tilde{d}^i)^2 \, .
\end{equation}
We then use the expression for the non-central $\chi^2(3)$-distribution, denoted as $P_3(\tilde{d}^2,\lambda^2)$, introducing the non-centrality parameter $\lambda$,
\begin{equation}
\lambda^2 =\sum_{i=\{x,y,z\}} \expect{\tilde{d}^i}^2 \, .
\end{equation}
In our case, we have $\lambda=d_\mathrm{kin}/\sigma_r$.
}

\ch{We now introduce the cumulative distribution function (CDF)
\begin{equation}
    C_3(\tilde{d},\lambda)
    = \int_{0}^{\tilde{d}^2} \mathrm{d}t P_3(t,\lambda^2)
    \, .
\end{equation}
so that
\begin{equation}
    1-C_3(\tilde{d},\lambda)=
    \int_{\tilde{d}^2}^\infty \mathrm{d}t P_3(t,\lambda^2)
    \, .
\end{equation}
Integrating this expression by part we find
\begin{equation}
   \sigma_r \int_0^\infty \mathrm{d}\tilde{d} (1-C_3(\tilde{d},\lambda)) =\sigma_r \int_0^\infty \sqrt{t}P_3(t,\la^2)dt  = \sigma_r \expect{\tilde{d}}=  \expect{d}\, .
    \label{eq:expect_d_1}
\end{equation}
}

\ch{
Using the expression for $P_3(\tilde{d}^2,\lambda^2)$ from~\cite{noncentral_chi2} allows to identify the cumulative distribution function with the $Q_M$-function defined in~\cite{Nuttall_QM}, Eq.~(1)
\begin{equation}
    1-C_3(\tilde{d},\lambda)=
    \int_{\tilde{d}^2}^\infty \mathrm{d}t P_3(t,\lambda^2)=
    Q_{3/2}(\lambda,\tilde{d})
    \, .
\end{equation}
Finally, using Equation~(10) of \cite{Nuttall_QM} to integrate the function $Q_{3/2}(\lambda,\tilde{d})$ and identifying the associated Laguerre polynomial $L^m_n$ with the confluent hypergeometric function of the first kind $_1F_1$, e.g.~in~\cite{confluent_hypergeometric}, we find the solution to the left hand side of \eqref{eq:expect_d_1} and obtain 
\begin{equation}
    \langle{d}\rangle = \sigma_r \sqrt{\frac{\pi}{2}} L_{1/2}^{1/2}
    \Big(- \frac{d_\mathrm{kin}^2}{2 \sigma_r^2}\Big) \, .
\end{equation}
}

\ch{
For the variance, we use the result of the mean of the $\chi^2$-function from e.g.~\cite{noncentral_chi2},
\begin{equation}
\expect{d^2}=\sigma_r^2 \expect{\tilde{d}^2} = d_\mathrm{kin}^2 + 3\sigma_r^2
\, .
\label{eq:expect_d_square}
\end{equation}
Equations~\eqref{eq:expect_d_1} and \eqref{eq:expect_d_square} result in Equation~\eqref{eq:mean_and_sigma_d}.
On a final note, the mean dipole amplitude can be approximated by $\expect{d}\approx \sqrt{ 8\sigma_r^2/\pi + d_\mathrm{kin}^2}$. This is motivated by the fact that setting $d_\mathrm{kin}=0$ the expected amplitude of the random dipole is $\sqrt{8\sigma_r^2/\pi}$.
The simple approach of $\expect{d} \approx \sqrt{\expect{d_r}^2+d_\mathrm{kin}^2}$ is surprisingly accurate for all $\sigma_r,d_\mathrm{kin}<0.1$ within $3\%$.
}
\section{Bulk flow velocity}
\label{app:bulk_flow}

The bulk flow velocity is the expected root mean square velocity of all observed sources.
If the sources are distributed accordingly to the normalized redshift distribution $n(z)$, one can define the window function
\begin{equation}
    W(\myvec{r}) = H\big(z(r)\big) \frac{n\big(z(r)\big)}{4\pi r^2}
    \, ,
\end{equation}
such that $\int \mathrm{d} x^3\,  W(\myvec{r}) = \int \mathrm{d}z \, n(z) = 1$.

The bulk flow of all the observed sources is expected to be  (in units of the speed of light)
\begin{equation}
\begin{aligned}
    \expect{\myvec{v}_\mathrm{flow}^2} &=
    \left\langle
    \Big( \int \mathrm{d}^3 x\ W(\myvec{r}) \myvec{v}(\myvec{r})\Big)^2
    \right\rangle =
    \\
    &=\frac{1}{2\pi^2} \integral{k}{}{} P_{\delta \delta}(k)
    \left( \integral{z}{}{} H(z) f(z) D_+(z) n(z)
    \frac{\sin \big(kr(z)\big)}{kr(z)}
    \right)^2 \, .
\end{aligned} \label{e:bulkV}
\end{equation}
We take the power spectrum $P_{\delta \delta}$ (in synchronous gauge) at zero redshift
from \class. The difference between using the non-linear or linear power spectrum is negligible because only the largest scales are  relevant.
For \eqref{e:bulkV} we
 use that the velocity has no curl, $\myvec{v}=\uvec{k}V$, and $\myvec{v}\cdot \myvec{k}=-f(z)H(z)D_+(z)\delta(0)$~\cite{bulk_flow_book} where $f$ is the growth function defined by  $f = \der{\ln D_+}{\ln a}$, which is well approximated by $f(z)\approx \Omega_m^{4/7}(z)$ in $\La$CDM~\cite{Hamilton_2001}.
For the redshift distributions of the two applications (see left panel of~\figref{fig:ang_power_spectrum_redshift}),
we find
\begin{align}
    \text{LSST/Euclid:}
    \qquad
    \expect{\myvec{v}_\mathrm{flow}^2} = \big(2.4 \times 10^{-5} \big)^2
\\
    \text{SKA:}
    \qquad
    \expect{\myvec{v}_\mathrm{flow}^2} = \big(2.0 \times 10^{-5} \big)^2 \, .
\end{align}
In our Universe, we expect a random velocity $v$ from a $\chi(3)$-distribution with $\expect{v^2} = \expect{\myvec{v}_\mathrm{flow}^2}$, pointing in a random direction~\cite{bulk_flow_book}.
This means that the mean rest frame of all observed sources and the rest frame defined by the CMB differ by $v$, which has a magnitude smaller than 2\% of the expected $\beta$.

\bibliographystyle{JHEP}
\bibliography{refs}

\end{document}